\documentclass[paper]{JHEP3} 

\usepackage{epsfig,multicol,multirow}
\usepackage{subfigure,latexsym}

\usepackage{graphics,amscd,amsfonts}
\usepackage{bm}
\usepackage{amsmath,amssymb}
\usepackage{comment}

\newcommand{\nt}{\nonumber\\}

\newcommand{\cN}{{\cal N}}

\newcommand{\ba}{\begin{eqnarray}}
\newcommand{\ea}{\end{eqnarray}}
\newcommand{\beq}{\begin{equation}}
\newcommand{\eeq}{\end{equation}}
\newcommand{\sss}[1]{\subsubsection*{#1}}
\newcommand{\back}{\!\!\!\!\!\!}

\newcommand{\hlam}{{\hat\lambda}}

\newcommand{\vth}{\vartheta}

\newcommand{\eq} {equation}
\newcommand{\eqa} {eqnarray}
\newcommand{\NN} {\mbox {$\nonumber$}}

\newcommand{\Slash}[1]{\ooalign{\hfil/\hfil\crcr$#1$}}

\allowdisplaybreaks

\title{
\bf\LARGE
Numerical studies of the ABJM theory for arbitrary $N$ 
at arbitrary coupling constant
}

\author{
Masanori Hanada,$^{a,b}$ Masazumi Honda,$^{b,c}$ 
Yoshinori Honma,$^{c}$ Jun Nishimura,$^{a,b,c}$ 
Shotaro Shiba$^{a,d}$ and Yutaka Yoshida$^a$
\vspace*{0.5cm} \\
\llap{$^a$}High Energy Accelerator Research Organization (KEK),\\
Tsukuba, Ibaraki 305-0801, Japan\\
\llap{$^b$}Kavli Institute for Theoretical Physics, University of California,\\
Santa Barbara, CA 93106-4030, USA\\
\llap{$^c$}Department of Particle and Nuclear Physics,\\
Graduate University for Advanced Studies (SOKENDAI),\\
Tsukuba, Ibaraki 305-0801, Japan\\
\llap{$^d$}Blackett Laboratory, Imperial College London,\\
London SW7 2AZ, UK
\vspace*{0.5cm} \\
\email{hanada@post.kek.jp, mhonda@post.kek.jp, yhonma@post.kek.jp,\\
jnishi@post.kek.jp, sshiba@post.kek.jp, yyoshida@post.kek.jp}}

\preprint{\footnotesize
KEK-TH-1516,\:NSF-KITP-12-026,\:Imperial/TP/2012/SS/01}

\abstract{
We show that the ABJM theory, which is an $\mathcal{N}=6$ superconformal
${\rm U}(N)\times {\rm U}(N)$ Chern-Simons gauge theory,
can be studied for arbitrary $N$ at arbitrary coupling constant
by applying a simple Monte Carlo method to the matrix model that
can be derived from the theory by using the localization technique.
This opens up the possibility of probing the quantum aspects of M-theory
and testing the $AdS_4/CFT_3$ duality at the quantum level. 
Here we calculate the free energy, 
and confirm the $N^{3/2}$ scaling in the M-theory limit 
predicted from the gravity side.
We also find that our results 
nicely interpolate
the analytical formulae proposed previously 
in the M-theory and type IIA regimes.
Furthermore,
we show that some results obtained by the Fermi gas approach
can be clearly understood from the constant map contribution obtained by the
genus expansion.
The method can be easily generalized 
to the calculations of BPS operators
and to other theories that reduce to matrix models.\footnote{The 
simulation code is available upon request to mhonda@post.kek.jp.}
}


\begin{document}

\setcounter{footnote}{0}

\section{Introduction}
Three-dimensional gauge theories with a Chern-Simons term 
have been studied extensively
for their rich and beautiful mathematical structure \cite{Witten:1988hf}
and for applications to quantum Hall systems in
condensed matter physics.
Their relevance to superstring/M-theory was realized
in 2004 \cite{Schwarz:2004yj}
by the appreciation that 
superconformal Chern-Simons theories coupled to 
matter fields capture the dynamics of M-theory 
on a multiple M2-brane background.
The latter theory is expected to be obtained
as an infrared fixed point of type IIA superstring theory 
on a D2-brane background,
whose low-energy description is 
given by the maximally supersymmetric $(2+1)$-dimensional Yang-Mills theory.   
In 2008, an explicit form of
the Chern-Simons theory that describes the infrared fixed point
was proposed by 
Aharony, Bergman, Jafferis and Maldacena (ABJM) \cite{Aharony:2008ug},  
following important earlier works \cite{Bagger:2007jr,Gustavsson:2007vu}. 
It is a $U(N)\times U(N)$ theory 
with Chern-Simons levels $k$ and $-k$ coupled to bifundamental matters. 
The on-shell supersymmetric Lagrangian of the theory is given by 
\begin{eqnarray}
&&\back\back
{\mathcal L}_{U(N)\times U(N)}
\nt
&&\back\back=
k\,\mathrm{Tr}\,\Biggl[
\frac12\epsilon^{\mu\nu\rho}\left(
-A_\mu\partial_\nu A_\rho-\frac{2}{3}A_\mu A_\nu A_\rho
+\tilde{A}_\mu\partial_\nu \tilde{A}_\rho
+\frac{2}{3}\tilde{A}_\mu \tilde{A}_\nu \tilde{A}_\rho
\right)
\nt&&\quad
+\left(-D_\mu\bar{\Phi}^\alpha D^\mu\Phi_\alpha
+i\bar{\Psi}^\alpha\Slash{D}\Psi_\alpha\right)
-i\epsilon^{\alpha\beta\gamma\delta}\Phi_\alpha\bar{\Psi}_\beta\Phi_\gamma\bar{\Psi}_\delta
+i\epsilon_{\alpha\beta\gamma\delta}\bar{\Phi}^\alpha\Psi^\beta\bar{\Phi}^\gamma\Psi^\delta
\nonumber \\
&&\quad
+i\left(
-\bar{\Psi}_\beta\Phi_\alpha\bar{\Phi}^\alpha\Psi^\beta
+\Psi_\beta\bar{\Phi}_\alpha\Phi^\alpha\bar{\Psi}^\beta
+2\bar{\Psi}_\alpha\Phi_\beta\bar{\Phi}^\alpha\Psi^\beta
-2\Psi^\beta\bar{\Phi}^\alpha\Phi_\beta\bar{\Psi}_\alpha
\right)
\nonumber\\
&&\quad
+\frac{1}{3}\left(
\Phi_\alpha\bar{\Phi}^\beta\Phi_\beta\bar{\Phi}^\gamma\Phi_\gamma\bar{\Phi}^\alpha
+\Phi_\alpha\bar{\Phi}^\alpha\Phi_\beta\bar{\Phi}^\beta\Phi_\gamma\bar{\Phi}^\gamma
\right. \nonumber \\
&~& \quad\quad \quad \left.
+4\Phi_\beta\bar{\Phi}^\alpha\Phi_\gamma\bar{\Phi}^\beta\Phi_\alpha\bar{\Phi}^\gamma
-6\Phi_\gamma\bar{\Phi}^\gamma\Phi_\beta\bar{\Phi}^\alpha\Phi_\alpha\bar{\Phi}^\beta
\right)
\Biggl] \,,
\label{PF-orgin}
\end{eqnarray}
where $A_\mu$ and $\tilde{A}_\mu$ are $U(N)$ gauge fields, and 
$\Phi_\alpha$ and 
$\Psi_\alpha\ (\alpha=1,2,3,4)$ are 
bosonic and fermionic complex bifundamental fields, respectively. 
This theory has ${\cal N}=8$ supersymmetry for $k=1,2$
and ${\cal N}=6$ supersymmetry for $k\ge 3$.
It has been conjectured to be dual to M-theory 
on $AdS_4\times S^7/{\mathbb Z}_k$ for $k\ll N^{1/5}$,
and to type IIA superstring on $AdS_4\times {\mathbb C}P^3$ 
in the planar large-$N$ limit with 
the 't Hooft coupling constant $\lambda=N/k$ kept fixed. 

From the viewpoint of quantum gravity, the ABJM theory
is important since it may provide us with a nonperturbative definition of 
type IIA superstring theory or M-theory on $AdS_4$ backgrounds 
since the theory is well-defined for finite $N$.
One may draw a precise analogy with the way maximally 
supersymmetric Yang-Mills theories may provide 
us with nonperturbative formulations 
of type IIA/IIB superstring theories 
on D-brane backgrounds through the gauge/gravity 
duality \cite{Maldacena:1997re,Gubser:1998bc,Witten:1998qj,Itzhaki:1998dd}.
In particular, the M-theory limit is important 
given that M-theory is not defined even perturbatively,  
although there is a well-known conjecture
on its nonperturbative formulation 
in the infinite momentum frame in terms of 
matrix quantum mechanics \cite{Banks:1996vh}. 
The planar limit, which corresponds to
type IIA superstring theory, has interest on its own
since it may allow us to perform more detailed tests of 
the gauge/gravity duality than in the case of $AdS_5/CFT_4$.
In particular, we may hope to calculate the $1/N$ corrections 
to the planar limit, which enables us to test the gauge/gravity duality
at the quantum string level, little of which is known so far.

In all these prospectives, one needs to study the ABJM theory
in the strong coupling regime.
As in the case of QCD, it would be nice if one could study 
the ABJM theory on a lattice by Monte Carlo methods.
This seems quite difficult, though, for the following three reasons.
Firstly, the construction of the Chern-Simons term on the lattice
is not straightforward, although there is a proposal 
\cite{Bietenholz:2000ca,Bietenholz:2003vw} based on its connection
to the parity anomaly.
Secondly, the Chern-Simons term is purely imaginary in the Euclidean
formulation, which causes a technical problem known
as the sign problem when one tries to apply the idea of importance
sampling.
Thirdly, the lattice discretization necessarily breaks supersymmetry,
and one needs to restore it in the continuum limit by fine-tuning
the coupling constants of the supersymmetry breaking relevant operators.
(See, for instance, ref.~\cite{Giedt:2009yd}.)
This might, however, be overcome by the use of 
a non-lattice regularization of the ABJM theory \cite{Hanada:2009hd}
based on the large-$N$ reduction on $S^3$ \cite{Ishii:2008ib,Kawai:2009vb},
which is shown to be useful in studying the planar limit of the 4d $\mathcal{N}=4$
super Yang-Mills theory \cite{Honda:2011qk,Nishimura:2009xm,Honda:2010nx}.

What we do here instead is to apply Monte Carlo methods not to the
original theory (\ref{PF-orgin}) but to a matrix model obtained after 
a huge reduction of the degrees of freedom due to supersymmetry.
In fact, 
it has been known for a while 
in certain supersymmetric theories
that one can reduce the path integral to a finite dimensional matrix model
by using the so-called localization technique.
Such a technique was applied \cite{Pestun:2007rz} to 
4d ${\cal N}=4$ super Yang-Mills theory,
and some conjecture on the half-BPS Wilson loops\footnote{This 
formula is also reproduced by a numerical simulation 
in the large-$N$ limit \cite{Honda:2011qk,Nishimura:2009xm}.
} \cite{Erickson:2000af,Drukker:2000rr} has been confirmed. 
In ref.~\cite{Kapustin:2009kz}, the same technique has been
applied to the ABJM theory on three-sphere $S^3$,
and its partition function was shown to reduce to a matrix integral 
\begin{eqnarray}
\label{PF2}
Z(N,k)
&=&
\frac{1}{(N!)^2}\int\frac{d^N\mu}{(2\pi)^N}\frac{d^N\nu}{(2\pi)^N} 
\nonumber \\
&~& \quad \quad \quad
\frac{\prod_{i<j}\Bigl(2\sinh\frac{\mu_i-\mu_j}{2}\Bigr)^2 
\Bigl(2\sinh\frac{\nu_i-\nu_j}{2}\Bigr)^2}
{\prod_{i,j} \Bigl(2\cosh\frac{\mu_i-\nu_i}{2}\Bigr)^2} 
\exp\left[\frac{ik}{4\pi}\sum_{i=1}^N (\mu_i^2-\nu_i^2)\right] \,, 
\end{eqnarray}
which is commonly referred to as the ABJM matrix model.\footnote{The 
localization of the ABJM theory 
and related theories on various manifolds such as 
$S^3$ \cite{sphere3}, $S^1 \times S^2$ \cite{SCFTindex} 
and squashed $S^3$ \cite{squashed} has also been considered.
}
By using this matrix model, 
the free energy of the ABJM theory has been studied intensively 
\cite{Marino:2009jd,Drukker:2010nc,Herzog:2010hf,Drukker:2011zy,
Marino:2011nm,Fuji:2011km,Okuyama:2011su,Marino:2011eh}. 
In ref.~\cite{Drukker:2010nc} the planar limit and the 
$1/N$ corrections\footnote{There is also
a study of non-planar corrections to anomalous dimensions by the integrability approach.
See e.g., \cite{Kristjansen:2008ib}.
} around it have been studied 
employing a technique from topological string theory,
and the on-shell action of 
the type IIA supergravity on $AdS_4 \times \mathbb{C}P^3$ has been reproduced. 
In ref.~\cite{Herzog:2010hf} 
the free energy in the M-theory limit has been obtained 
using some ansatz for the eigenvalue distribution. 
In ref.~\cite{Fuji:2011km} the genus expansion at
strong 't Hooft coupling has been considered and a resummed form
was obtained in terms of the Airy function   
by using the holomorphic anomaly equation \cite{HAE}.
The obtained simple form was claimed to be valid to all orders in the genus
expansion up to the worldsheet instanton effect.
In ref.~\cite{Marino:2011eh}, the free energy
in the M-theory regime at small $k$ has been calculated by the
Fermi gas approach, and the result turns out to be given by
the Airy function obtained in ref.~\cite{Fuji:2011km} with 
some extra terms.
These results, if correct, would enable us 
to shed light on the dynamical aspects of M-theory
and to test
the AdS/CFT duality 
including the string loop effect
by studying the gravity side further.

In this paper we show that
the ABJM matrix model can be rewritten
in a form suitable 
for Monte Carlo simulations, which enables
simple calculation of the partition function and BPS operators
for arbitrary values of the rank $N$ and the level $k$ from first
principles.
In particular, we calculate the partition function explicitly
for various $N$ and $k$,
which is supposed to contain
the nonperturbative effects
corresponding to the worldsheet instantons in string theory
neglected in refs.~\cite{Drukker:2010nc,Fuji:2011km}.
We find the well-known constant map contribution \cite{HAE,constant_map,Marino:2009dp} is also correctly reproduced.
The constant map contributions
are of the order of $\lambda^0/k^{2g-2}$, where $g$ is the genus.
Note that these terms depend on 
the string coupling\footnote{This is the string coupling 
in the context of topological string theory. 
The string coupling in the IIA superstring limit of the ABJM theory 
is $g_{s,\rm IIA}^2 \sim \sqrt{\lambda}/k^2\sim\sqrt{\lambda}g_s^2$.} 
$g_{s}=2\pi i/k$ but not on the 't Hooft coupling constant.
In the planar limit, they correspond to a constant shift of the free energy, 
which becomes negligibly small in the large 't Hooft coupling limit,
and hence they do not spoil the agreement with the type IIA supergravity.
They are negligible in the eleven-dimensional supergravity limit as well. 
However, when one considers quantum strings or M-theory, 
these terms should be taken into account
since they have an explicit $g_s$ dependence.  
In fact, these terms become dominant at strong 't Hooft coupling for $g\ge 2$.

This paper is organized as follows. 
In section \ref{sec:F-lim} we review the previous results for
the free energy 
of the ABJM matrix model 
in various limits, which are obtained by analytical methods.
In section \ref{sec:methods} we describe our numerical method.
In section \ref{sec:FE} we present our results, and discuss the
discrepancies from the analytical results.
In section \ref{sec:deviation_origin}
we show that these discrepancies can be interpreted 
as the constant map contributions.
Section \ref{sec:con} is devoted to a summary and discussions.
In Appendix \ref{appendix:simulation} we explain the
basics and some details of Monte Carlo simulation.
In Appendix \ref{appendix:derivation} we derive
an alternative form of the ABJM matrix model, which is suitable for 
Monte Carlo simulation.
In appendix \ref{appendix:con_fermi} we show the equivalence 
between the constant map contribution and 
the Fermi gas result $A(k)-\frac{1}{2}\log{2}$,
which is derived by the large-$k$ and small-$k$ expansions, respectively.

\paragraph{Note added.}
After the first version appeared on the arXiv, we were informed by
Marcos Mari\~no that
the discrepancies observed at genus 0, 1 and 2 
between our numerical results 
and the formula proposed in ref.\,\cite{Fuji:2011km}
can be naturally attributed to the constant map contributions.
We greatly appreciate this important comment, which 
enabled us to deepen our argument in 
section \ref{sec:constant_map_vs_Fermi_gas}.

\section{Previous analytical results for the free energy}
\label{sec:F-lim}
%
In this section we summarize some known analytical results 
for free energy of the ABJM theory, which is defined 
in terms of the partition function (\ref{PF2}) as
\ba
F(N,k)\,=\, \log Z(N,k) \,.
\label{defF}
\ea

\subsection{Perturbative results for all $N$}
The free energy can be calculated by using a usual perturbative technique, 
and the result at the one-loop level is given as 
(See, for example, ref.~\cite{Marino:2011nm}.)
\ba\label{finNsL}
F_{\rm weak}
\!\!&=&\!\!
-N^2\log\frac{2N}{\pi\lambda}-N\log 2\pi+2\log G_2(N+1)
\\
&\stackrel{N\gg 1}{=}&
N^2\left(\log 2\pi\lambda-\frac32-2\log 2\right)
-\frac16\log N+2\zeta'(-1)
+\sum_{g=2}^\infty \frac{B_{2g}}{g(2g-2)}N^{2-2g} \,,
\label{finNsL2}
\ea
where $G_2(x)$ is the Barnes G-function $G_2 (x)\equiv
\prod_{s=1}^{x-2} s!$. 
The $1/N$-expansion is shown in 
the second line
with Riemann's zeta function $\zeta(x)$
and the Bernoulli numbers $B_{2g}$.
The O($N^2$) terms in (\ref{finNsL2}) 
agree with the result (\ref{weak}) obtained 
in the planar limit.
Note, however, that the expression (\ref{finNsL}) includes 
contributions to all orders in the $1/N$-expansion. 

\subsection{$N=2$ with arbitrary $k$}
An exact expression for $N=2$ is obtained 
by Okuyama \cite{Okuyama:2011su} 
as\footnote{Note that the normalization of the 
partition function adopted in ref.~\cite{Okuyama:2011su} 
differs from ours as $Z_{\rm{Okuyama}} = 2^{2N} Z_{\rm{ours}}$.
}
\ba\label{exactN2}
F (2,k)\,=\,\begin{cases}
\,\displaystyle{
\log\left[\frac1{k}\sum_{s=1}^{k-1}(-1)^{s-1}
\left(\frac12-\frac{s}{k}\right)\tan^2\frac{\pi s}{k}
+\frac{(-1)^\frac{k-1}{2}}{\pi}\right] -4\log{2}}
&\text{for odd $k$}\\[+14pt]
\,\displaystyle{
\log\left[\frac1{k}\sum_{s=1}^{k-1}(-1)^{s-1}
\left(\frac12-\frac{s}{k}\right)^2\tan^2\frac{\pi s}{k}\right] -4\log{2}}
&\text{for even $k$} \,.
\end{cases}
\ea

This result has been obtained by direct integration of (\ref{grand_cano}).
Since the expressions for the odd and even $k$ cases are different,
the analyticity in $k$ (when one regards $k$ or equivalently 
the 't Hooft coupling constant $\lambda$ 
as a continuous variable) is not obvious a priori.
However, as we will see in section \ref{sec:FE},
our numerical results suggest that the free energy is a 
smooth function of $k$.
The analyticity is important in the context of the AdS/CFT
correspondence, in which one assumes the analyticity on the gravity side.
Also the analysis in the planar limit usually assumes the
analyticity implicitly.
%
%
\subsection{Planar limit ($N\to\infty$ with $\lambda $ \rm{fixed})}
The free energy in the planar limit ($N\to\infty$ with $\lambda $ \rm{fixed})
has been calculated by 
Drukker, Marino and Putrov (DMP) \cite{Drukker:2010nc}.
These results have been obtained by a standard
matrix model technique after the analytic 
continuation \cite{Marino:2009jd} to the lens space $L(2,1)=S^3 /\mathbb{Z}_2$ 
matrix model \cite{Marino:2002fk,Aganagic:2002wv}, 
which is obtained from the pure Chern-Simons theory on $L(2,1)$.
The validity of the analytic continuation is proved 
diagramatically in refs.~\cite{Dijkgraaf:2003xk,Dijkgraaf:2002pp}.

At weak coupling ($\lambda\ll 1$) the authors obtain
\ba\label{weak}
F_{\rm{weak,planar}} = N^2\left(\log 2\pi\lambda-\frac32-2\log 2\right)
\ea
up to ${\rm O}(\lambda )$.

At strong coupling ($\lambda\gg 1$) the authors obtain
\ba
\label{strong}
F_{\rm DMP}=
-\frac{\pi\sqrt{2}}{3}\frac{\hat{\lambda}^{3/2}}{\lambda^2}N^2 
\qquad\quad
\mbox{where}\quad \hat{\lambda}=\lambda -\frac{1}{24}
\ea
to all orders of the $1/\lambda$ expansion.
The leading behavior 
$F_{\rm DMP}\simeq -\sqrt{2}\pi N^2 /(3\sqrt{\lambda})$ 
agrees with the dual type IIA supergravity 
prediction \cite{Drukker:2010nc,Emparan:1999pm}
including the overall coefficient.
It has been claimed that the free energy (\ref{strong}) 
at strong coupling receives the correction of the form 
\[
\simeq \frac{N^2}{\lambda^2} \sum_{l\geq 1} 
e^{-2\pi l \sqrt{2\hat{\lambda}} } f_I^{(l)} \left(  
\frac{1}{\pi\sqrt{2\hat{\lambda}}} \right) \,,
\]
where $f_I^{(l)}(x)$ is a polynomial in $x$ of degree $2l-3$ (for $l\geq 2$).
This exponentially small correction has been interpreted in 
ref.~\cite{Drukker:2010nc}
as the effect 
of the worldsheet instanton in the dual type IIA superstring, 
which corresponds to a string worldsheet wrapping a $\mathbb{C}P^1$ cycle 
in $\mathbb{C}P^3$ \cite{Cagnazzo:2009zh}.

In section \ref{sec:FE} we will show that 
another contribution of the order of O($N^2/\lambda^2$)
due to the constant map
needs to be added in comparing with precise numerical analysis.
Although this term does not affect the agreement with supergravity, 
it must be taken into account when one compares the finite $\lambda$
corrections with the string $\alpha'$ corrections.

\subsection{M-theory limit ($N\to\infty$ with $k $ \rm{fixed})}
In ref.~\cite{Herzog:2010hf}, the free energy 
in the M-theory limit\footnote{Strictly speaking,
since the ABJM theory has been conjectured to be dual to the 
M-theory for $k\ll N^{1/5}$,
the limit $N\to\infty$ with $k$ \rm{fixed} is merely a sufficient condition.
In the following, however, we simply call it ``the M-theory limit''.
} ($N\to\infty$ with $k $ \rm{fixed}) 
has been calculated and confirmed the prediction
\ba
F_{\rm SUGRA}=-\frac{\pi\sqrt{2k}}{3} N^{3/2} 
\label{M-N3-2}
\ea
from the dual eleven-dimensional supergravity,
which shows the well-known $N^{3/2}$ scaling 
for the degrees of freedom in the theory of 
M2-branes \cite{Klebanov:1996un}.
Note also that (\ref{M-N3-2})
agrees with what one obtains formally 
from the leading large-$\lambda$ behavior of the planar
result (\ref{strong}) by replacing $\lambda$ with $N/k$.

The result (\ref{M-N3-2}) was obtained by imposing 
an ansatz for the eigenvalue distribution
\[
\mu_i = N^\alpha z_i +iw_i \,,\quad  \nu_i = 
N^\alpha z_i -iw_i \quad (z_i ,w_i \in \mathbb{R}) \,,
\]
which is necessary for the cancellation of long-range forces,
and is also suggested by numerical studies of the saddle point equation. 
The parameter $\alpha$ is chosen to be $1/2$ by 
requiring
that all the short-range forces contribute to the free energy at the same order of $N$
 in order to have nontrivial solutions.

\subsection{$1/N$ expansion around the planar limit}

Fuji, Hirano and Moriyama (FHM) \cite{Fuji:2011km}
studied the free energy to all orders in the genus expansion
neglecting the instanton contribution, which is of the order of
${\rm O}(e^{-2\pi \sqrt{\lambda}})$.
Their proposal for a resummed form is given by
\begin{equation}
F_{\rm FHM}(N,\lambda ) \,=\, \log \left[
\frac{1}{\sqrt{2}}  \left( \frac{4\pi^2 N}{\lambda} \right)^{1/3} 
\mathrm{Ai}\left[\left(\frac{\pi}{\sqrt{2}} \left(\frac{N}{\lambda}\right)^2
\lambda_{\rm ren}^{3/2}\right)^{2/3} \right]\right] \,, 
\label{FHM}
\end{equation}
where $\mathrm{Ai}(x)$ is the Airy function, 
and the ``renormalized 't Hooft coupling'' $\lambda_{\rm ren}$ is given by  
\begin{equation}
\label{lambda_ren}
\lambda_{\rm ren}=\lambda-\frac{1}{24} -\frac{\lambda^2}{3N^2} \,. 
\end{equation}
The appearance of the Airy function \cite{Fuji:2011km}
is also encountered in the context of M-theory flux 
compactification \cite{Ooguri:2005vr}.
Note that the expression (\ref{FHM}) reproduces (\ref{strong}) 
in the large-$N$ limit as one can easily see by using 
the asymptotic formula 
$\log{\rm{Ai}(x)} \sim -\frac{2x^{3/2}}{3}$ for $x \gg 1$.
In section \ref{sec:FE} we will show that (\ref{FHM}) 
has
another contribution, 
which is necessary for comparison with our numerical results. 

The free energy at higher genus 
has been studied earlier \cite{Drukker:2010nc,Drukker:2011zy}
by using a topological string technique
after analytic continuation to the lens space matrix model.
The analysis in ref.~\cite{Fuji:2011km}
has been performed by using the holomorphic anomaly equation \cite{HAE},
whose solution is the same as the one 
for the loop equation \cite{Ambjorn:1992gw,Akemann:1996zr} 
with some appropriate boundary conditions.
In order to solve the holomorphic anomaly equation,
one needs to provide some inputs
such as the free energy at genus zero and one,
which are taken to be
\[
F_{\rm{FHM}}^{(0)} = \frac{4\sqrt{2}\pi^3}{3} 
\hat{\lambda}^{3/2} \quad {\rm and} \quad 
F_{\rm{FHM}}^{(1)} = \frac{\pi}{3\sqrt{2}}
\hat{\lambda}^{1/2} -\frac{1}{4}\log{(8\hat{\lambda})} \,.
\] 
In this way the authors have found a general solution,
which gives the free energy at all genus 
up to the worldsheet instanton effect.
The integration constants were determined by assuming 
the absence of non-perturbative corrections 
of the type $\sim O\left( e^{-1/g_s^2}\right)$.
Strictly speaking, what one obtains in this way is
the ``weight zero'' contribution to the free energy in the language of
topological string theory. It is claimed that one can turn 
this result into
the one including contributions from all weights by
making a replacement 
$\lambda \rightarrow \lambda_{\rm{ren}}$,
which is given in (\ref{lambda_ren}).

This ``renormalized 't Hooft coupling'' is different from 
the expectation from the gravity side \cite{Bergman:2009zh}\,:
$\lambda_{\rm{ren,grav}}=\lambda -1/24 +\lambda^2 /(24N^2 )$.
While it is possible that this disagreement may imply that 
the AdS/CFT does not hold at finite-$N$/quantum string level, 
we should definitely gain more understanding on both gauge theory 
and gravity sides.
The additional contribution to  
the FHM result from the constant map should be
important also from this point of view.

\subsection{$N\gg 1 $, small $k$}
\label{sec_fermi}
In ref.~\cite{Marino:2011eh}, the free energy 
with fixed small $k$ has been calculated 
by using the Fermi gas approach 
neglecting the quantum mechanical instanton effect (worldsheet instanton)
and the terms which are suppressed exponentially at large $N$ (membrane instanton).
In this approach,
the partition function of the ABJM theory is 
regarded as an ideal Fermi gas system described by (\ref{fermi_wave}) 
with the Planck constant identified as $\hbar =2\pi k$.
The result is given by
\begin{eqnarray}
F_{\rm{Fermi}}
= \log \left[
\frac{\left( 4\pi^2 k \right)^{1/3}}{\sqrt{2}}   
\mathrm{Ai}\left[ \left(\frac{\pi k^2}{\sqrt{2}} \right)^{2/3}
\left( \frac{N}{k} -\frac{1}{24} -\frac{1}{3k^2} \right)
\right]\right]  
+A(k)-\frac{1}{2}\log{2} \,.
\label{marino-largeN-smallk}
\end{eqnarray}
The leading large-$N$ behavior reproduces eq.\,(\ref{M-N3-2}) exactly. 
The function $A(k)$ in (\ref{marino-largeN-smallk}) 
is given for $k\ll 1/(2\pi )$ as\footnote{Although 
the Chern-Simons level $k$ must be integer in a physical setup, 
the integral \eqref{PF2} is itself 
well-defined also for non-integer $k$ and
we can actually obtain numerical results, which turn out to be 
a smooth function of $k$.}
\ba
A(k) =
\frac{2\zeta (3)}{\pi^2 k}  
-\frac{k}{12} 
-\frac{\pi^2 k^3}{4320} +{\rm O}(k^5)\,.
\label{Ak-expand}
\ea

Since the first term in (\ref{marino-largeN-smallk}) can be obtained formally 
from the FHM result $F_{\rm{FHM}}$ in (\ref{FHM})
by replacing $\lambda$ with $N/k$,
one can rewrite it as
\begin{eqnarray}
\label{fermi}
F_{\rm{Fermi}}
= F_{\rm FHM}+A(k)-\frac{1}{2}\log{2} \,,
\end{eqnarray}
where $A(k)$ may be viewed as 
``quantum corrections'' with the ``Planck constant'' 
$\hbar =2\pi k$. 
Note that
the first term in (\ref{fermi}) is valid for all $k$
although (\ref{Ak-expand}) is obtained at small $k$.
The authors note that the second and third terms 
in (\ref{Ak-expand}) are given by
\ba\label{Ak1}
A(k) 
&=&
\frac{2\zeta (3)}{\pi^2 k}  
-\sum_{n=1}^\infty \frac{(-1)^{n-1}B_{2n}}{n(2n-1)(2n)!} \pi^{2n-2}k^{2n-1}
\nt&=&
\frac{2\zeta (3)}{\pi^2 k}  
-\frac{2}{\pi} \int_0^{\pi k} \frac{d\xi}{\xi^2}\, 
\log\left[  \frac{\sin(\xi/2)}{\xi/2} \right]
\ea
for $n=1,2$.
This is the power series with
odd powers of $k$
unlike the usual genus expansion 
around the planar limit. 
The authors suggest that 
$A(k)$ may encode the effect from D0-branes of the order of
${\rm O}\left( e^{-k} \right) \sim 
{\rm O} \left( e^{-1/g_s} \right)$.

Since this analysis for $A(k)$ assumes $k\ll 1/(2\pi )$,
it is not clear a priori whether the result holds
at physical values of $k$ corresponding to integers.
As we will see later, (\ref{Ak-expand}) and 
(\ref{Ak1}) are in reasonable agreement with 
our numerical result for small $k$ such as $k=1,2,3$, 
but not for larger $k$ (including the planar limit).

\section{Numerical methods for the ABJM matrix model at
arbitrary $N$ and $k$}
\label{sec:methods}
In this section we discuss how we can study 
the ABJM matrix model at arbitrary $N$ and $k$
by applying a standard Monte Carlo method.
For the readers who are not familiar with Monte Carlo methods
in general, we review the basic ideas in 
Appendix \ref{appendix:simulation}.
For an earlier work on Monte Carlo simulation of a one-matrix model,
see ref.~\cite{Kawahara:2007eu}.

The ABJM matrix model in the form \eqref{PF2} is not suitable 
for Monte Carlo simulation 
since the integrand is not real positive.\footnote{One might 
think of simulating a system without the phase factor
$\exp((ik/4\pi)(\mu_i^2-\nu_i^2))$, and including its effect
afterwards by reweighting.
While it is possible to obtain results for the $k=1$ case 
along this line, the calculation becomes more and more difficult
for larger $k$ due to the sign problem.
} 
However, as we review in Appendix~\ref{appendix:derivation}
in detail, one can rewrite the ABJM matrix model as follows.
\begin{eqnarray}
Z(N,k)&=&C_{N,k}\,g(N,k) \,, \quad 
C_{N,k}=\frac{1}{(4\pi k)^N\,N!}  \,, \nonumber \\
g(N,k)&=&
\int d^Nx\frac{\prod_{i<j}\tanh^2{\left( \frac{x_i-x_j}{2k}\right)}}
{\prod_i 2\cosh(x_i/2)} \,. 
\label{action_sign_free_1}
\end{eqnarray}
In the $k=1$ case, 
one may view (\ref{action_sign_free_1}) 
as a mirror description of the ABJM theory
in terms of 
the 3d $U(N)$ $\mathcal{N}=4$ SYM 
with adjoint and fundamental hypermultiplets, which is
isomorphic to 3d $U(N)$ $\mathcal{N}=8$ SYM
in the low-energy limit \cite{Kapustin:2010xq}.
The important point here is that,
in this form (\ref{action_sign_free_1}), 
the integrand is real positive, and we can perform 
Monte Carlo simulation in a straightforward manner
as described in Appendix \ref{appendix:simulation}.

We should also note that, 
while the level $k$ should be an integer in the original 3d gauge theory, 
nothing prevents us from considering non-integer $k$
in the integral \eqref{PF2}. 
In what follows, we therefore
extend the value of $k$ to any real number.

In order to calculate the free energy (\ref{defF}), 
which is the log of the partition function, we need to rewrite it
in terms of expectation values of some quantities, 
which are directly calculable by Monte Carlo methods.\footnote{For
applications of such an idea on different supersymmetric systems, 
see refs.~\cite{Krauth:1998xh} and  \cite{Kanamori:2010gw}.}
The basic idea in our case
is to calculate the ratios of the partition functions
for different $k$ or $N$ as expectation values.
Since we know the results for $k=0$ or $N=1$, we can obtain
results for arbitrary $k$ and $N$ by calculating
an appropriate product of the ratios.
Depending on whether we change $k$ or $N$, we have the following two
methods, which give the same result
within statistical errors as we have checked for various $k$ and $N$. 
The second method is particularly useful in studying the M-theory limit,
which corresponds to the large $N$ limit with fixed $k$.

\subsection{Calculating the ratio of partition functions with different $k$}
Let us consider a trivial identity
\begin{eqnarray}
\frac{g(N,k_2)}{g(N,k_1)}
=
\frac{\int d^Nx\,e^{-S(N,k_2;x)}}{\int d^Nx\,e^{-S(N,k_1;x)}}
=
\left\langle
e^{-S(N,k_2;x)+S(N,k_1;x)}
\right\rangle_{N,k_1} \,,
\label{ratio-k1k2}
\end{eqnarray}
where we have defined
\begin{eqnarray}
e^{-S(N,k;x)}
= \frac{\prod_{i<j}\tanh^2(\frac{x_i-x_j}{2k})}{\prod_i 2\cosh(x_i/2)}
\end{eqnarray}
and $\langle\,\cdots\,\rangle_{N,k}$ stands 
for the expectation value with respect to the action $S(N,k;x)$
\begin{eqnarray}
\langle {\cal O}\rangle_{N,k}
= \frac{\int d^Nx\,{\cal O}(x)e^{-S(N,k;x)}}{\int d^Nx\,
  e^{-S(N,k;x)}} \,.
\end{eqnarray}
The quantity (\ref{ratio-k1k2}) can be calculated easily
by the standard Monte Carlo method as far as $k_1$ and $k_2$
are sufficiently close.\footnote{As $k_2$ moves away from $k_1$,
the quantity $e^{-S(N,k_2;x)+S(N,k_1;x)}$ fluctuate violently 
during the simulation of the system $S(N,k_1;x)$,
which leads to larger statistical errors.
} 
Therefore, we can calculate the free energy $F$ as 
\begin{eqnarray}
F
&=& \log Z=\log C_{N,k}+\log g(N,k) \nonumber \\
&=&
\log C_{N,k}
+\sum_{i=1}^l \log \frac{g(N,k_i)}{g(N,k_{i-1})}
+\log g(N,0)
\nonumber\\
&=&
\log C_{N,k}
+\sum_{i=1}^l \log 
\left\langle
e^{-S(N,k_i;x)+S(N,k_{i-1};x)}
\right\rangle_{N,k_{i-1}}
+N\log\pi  \,,
\end{eqnarray}
where $0=k_0<k_1<\cdots<k_l=k$ and we have used
$g(N,0)= \int \frac{d^Nx}{\prod_i 2\cosh(x_i/2)}= \pi^N $
in the last line. 
We have to make the adjacent values of $k$ close enough
for the reason mentioned above.

\subsection{Calculating the ratio of partition functions with different $N$}
Let us decompose $N$ into $N=N_1+N_2$ and consider the ratio
\begin{eqnarray}
\frac{g(N,k)}{g(N_1,k)g(N_2,k)} 
&=&
\frac{\int d^N x\, e^{-S(N,k)} }
{\int d^N x\, e^{-S(N_1,k;x_1,\cdots,x_{N_1})-S(N_2,k;x_{N_1+1},\cdots,x_N)}}
\nonumber\\
&=&
 \left\langle 
e^{S(N_1,k;x_1,\cdots,x_{N_1})+S(N_2,k;x_{N_1+1},\cdots,x_N)-S(N,k)}
\right\rangle_{N_1,N_2} \,,
\label{reweight}
\end{eqnarray}
where the symbol $\langle \cdots \rangle_{N_1,N_2}$ 
denotes the expectation value with respect to
the ``action'' $S(N_1,k;x_1,\cdots,x_{N_1})+S(N_2,k;x_{N_1+1},\cdots,x_N)$.
Note that 
\begin{equation}
e^{S(N_1,k;x_1,\cdots,x_{N_1})+S(N_2,k;x_{N_1+1},\cdots,x_N)-S(N,k)} 
\,=\, \prod_{i=1}^{N_1} \prod_{J=N_1 +1}^N
              \tanh^2{\left( \frac{x_i -x_J}{2k} \right)} \,,
\label{crucial-rel}
\end{equation}
due to the factorization of the potential terms.
In order to calculate the right-hand side of (\ref{reweight})
with good accuracy, 
it is necessary to take $N_2$ small enough to make sure that 
(\ref{crucial-rel}) does not 
fluctuate violently during the simulation.  
In actual calculation we use $N_2 =1$.
Then by calculating (\ref{reweight}) for $N_1 = 1,2,3,\cdots$,
and by using the $N=1$ result
\begin{eqnarray}
g(1,k)=\int \frac{dx}{2\cosh(x /2)} = \pi  \,,
\end{eqnarray}
we can calculate the free energy for $N=2,3,4,\cdots$ successively
with a fixed value of $k$.

\section{Results for the free energy}
\label{sec:FE}

\begin{figure}[t]
\begin{center}
\includegraphics[width=7.4cm]{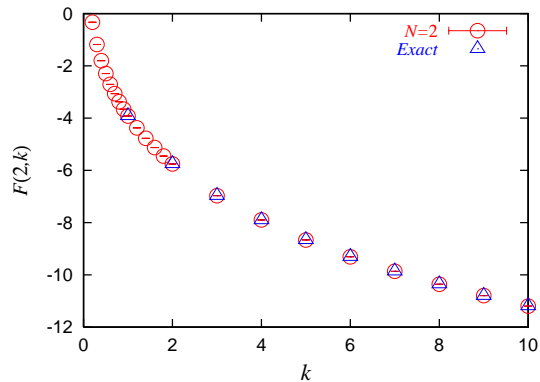}
\end{center}
\caption{The free energy of the 
ABJM theory for $N=2$ is plotted against the Chern-Simons level $k$.
The circles and triangles represent
our Monte Carlo result
and the exact result (\ref{exactN2}), respectively.
}
\label{fig:U(2)_numerics_vs_Okuyama}
\end{figure}

In this section we present our numerical result 
for the free energy of the ABJM theory.
In order to test our code, we first study 
the $N=2$ case and compare 
our result against the exact result (\ref{exactN2}) 
obtained by Okuyama \cite{Okuyama:2011su}.
As can be seen from fig.\,\ref{fig:U(2)_numerics_vs_Okuyama},
our result reproduces the exact result very accurately.
We have also obtained results for non-integer values of $k$,
which are not obtained in ref.~\cite{Okuyama:2011su}.
They are found to connect the results for integer $k$ smoothly.

\subsection{Planar limit}
\begin{figure}[t]
\begin{center}
\includegraphics[width=7.4cm]{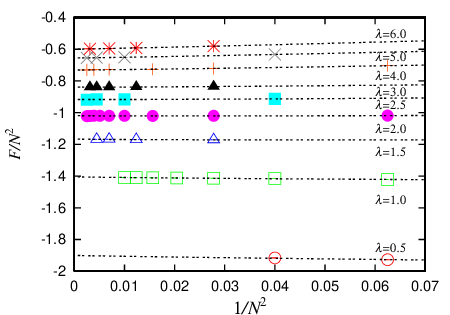} 
\includegraphics[width=7.4cm]{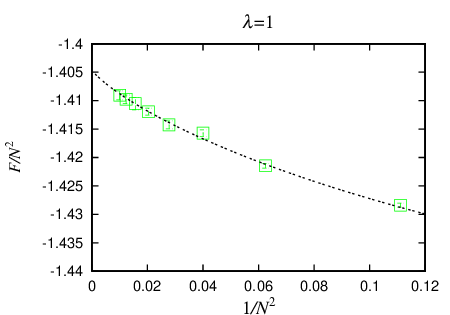} 
\end{center}
\caption{
The normalized free energy $F/N^2$ is plotted against $1/N^2$
for various values of $\lambda$ (Left). In the right panel,
we zoom up the plot for $\lambda=1$.
The data can be nicely fitted to 
$F(N,\lambda)/N^2=f_0(\lambda)+f_1(\lambda)/N^2 -\frac{1}{6}\log{N}$,
which enables us to make
a reliable extrapolation to the planar $N\rightarrow
\infty$
limit.
}
\label{fig:extrapolation_planar}
\end{figure}

\begin{figure}[t]
\begin{center} 
\includegraphics[width=7.4cm]{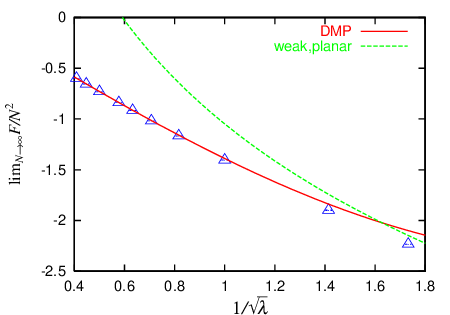}
\includegraphics[width=7.4cm]{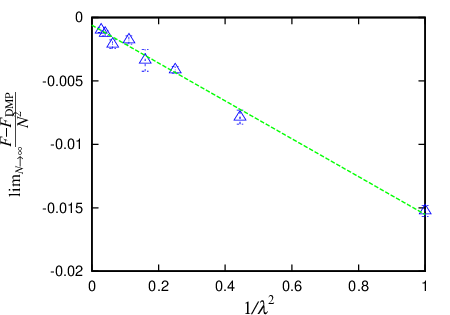} 
\end{center}
\caption{
(Left) The free energy in the planar limit
$f_0(\lambda)=\lim_{N\to\infty} F(N,\lambda)/N^2$ 
extracted from fig.~\ref{fig:extrapolation_planar}
is plotted against $1/\sqrt{\lambda}$.
Our results seem to interpolate the DMP result 
at strong coupling and the perturbative result at weak coupling.  
(Right) The difference between our result
and the DMP result, i.e., $\lim_{N\rightarrow\infty}(F-F_{\rm DMP})/N^2$,
is plotted against $1/\lambda^2$.
The data points can be fitted to a straight line,
which implies (\ref{d_planar}) and (\ref{d_planar2}).
}
\label{fig:free_energy_planar}
\end{figure}

Next we consider the planar limit ($N\to\infty$ with $\lambda=N/k$ fixed), 
which is conjectured to be dual to the 
classical type IIA superstring on $AdS_4\times {\mathbb C}P^3$. 
In fig.~\ref{fig:extrapolation_planar} we plot 
the normalized free energy $F/N^2$ against $1/N^2$
for various values of $\lambda$. 
Our results can be fitted well by
$F(N,\lambda)/N^2=f_0(\lambda)+f_1(\lambda)/N^2 -\frac{1}{6}\log{N}$
as theoretically
expected\footnote{The
functions $f_0(\lambda)$ and $f_1(\lambda)$ defined here are related to 
$F_0(\lambda)$ and $F_1(\lambda)$, which are defined in 
(\ref{genus}), as $f_0(\lambda)=-F_0(\lambda)/4\pi^2\lambda^2$ and
$f_1(\lambda)=F_1(\lambda)$.
}. 
In the left panel of fig.~\ref{fig:free_energy_planar}, 
we plot $f_0 (\lambda) =
\lim_{N\to\infty} F(N,\lambda)/N^2$ against $1/\sqrt{\lambda}$.
The results seem to interpolate 
the DMP result (\ref{strong}) at strong coupling
and the perturbative result (\ref{weak}) at weak coupling.  
However, by looking more carefully into the asymptotic behavior
for large $\lambda$, we find
certain discrepancies.
This can be seen from 
the right panel of fig.~\ref{fig:free_energy_planar},
in which we plot the difference $\lim_{N\rightarrow\infty}(F-F_{\rm
  DMP})/N^2$,
which is found to behave as
\begin{eqnarray}
\label{d_planar}
&~& \lim_{N\rightarrow\infty} \frac{F-F_{\rm DMP}}{N^2}
\stackrel{\lambda \gg 1}{\simeq} 
 \frac{a_0}{\lambda^2}+b_0  \,,   \\
&~& a_0 = -0.015      \pm 0.001   \,,\quad
b_0 = -0.0006  \pm 0.0002  
\label{d_planar2}
\end{eqnarray}
instead of the behavior 
${\rm O}(e^{-2\pi \sqrt{\lambda}})$ expected from 
the worldsheet instanton effect. 
We consider that $b_0$ is consistent with zero
since the fitting error may well be slightly underestimated.
Since the discrepancy (\ref{d_planar}) 
vanishes at $\lambda=\infty$ 
(assuming that $b_0$ in (\ref{d_planar}) is zero),
it does not affect the agreement with the dual type IIA supergravity.

In section \ref{sec:deviation_origin}
we explain that this discrepancy can be understood as the constant map at genus $0$. 
Similar discrepancies exist also
in $1/N$ corrections around the planar limit as we will see.

\subsection{M-theory limit}

\begin{figure}[t]
\begin{center} 
\includegraphics[width=7.4cm]{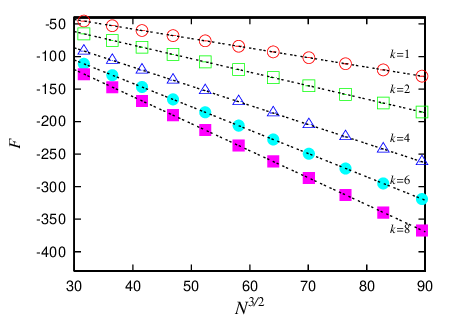} 
\includegraphics[width=7.4cm]{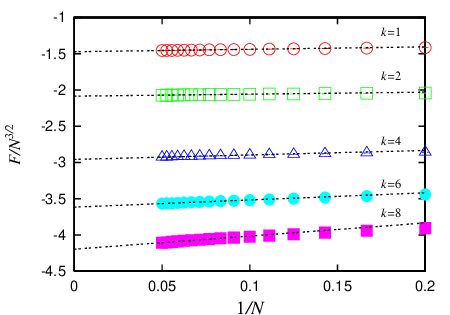} 
\end{center}
\caption{
(Left) The free energy is plotted against $N^{3/2}$ for
$k=1,2,4,6,8$. The data points can be fitted to straight lines,
which implies $F \sim N^{3/2}$ as $N$ increases.
(Right) The normalized free energy $F/N^{3/2}$ 
is plotted against $1/N$.
The data can be nicely fitted to straight lines,
which enables us to make extrapolations to the M-theory limit
reliably.
}
\label{fig:M_extrapolation}
\end{figure}

Next we consider the large-$N$ limit with fixed $k$, 
which is conjectured to correspond to the eleven dimensional supergravity 
on $AdS_4\times S^7/{\mathbb Z}_k$. 
Figure~\ref{fig:M_extrapolation} shows that
the free energy $F$ grows in magnitude as $N^{3/2}$ with $N$,
and $F/N^{3/2}$ behaves as 
$F(N,k)/N^{3/2}=h_0 (k)+h_1 (k)/N$, 
which enables us to obtain the M-theory limit
$h_0 (k)=\lim_{N\to\infty}F(N,k)/N^{3/2}$ reliably.

\begin{figure}[t]
\begin{center} 
\includegraphics[width=7.4cm]{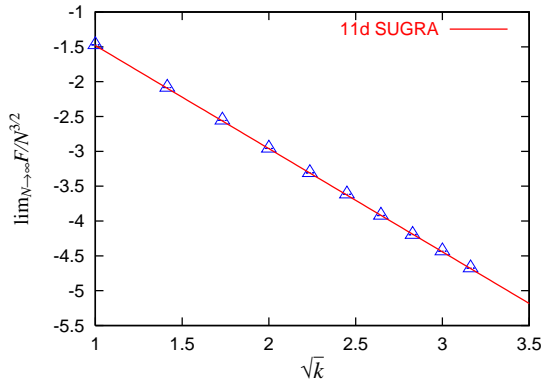} 
\end{center}
\caption{
The M-theory limit of the free energy
$\lim_{N\rightarrow\infty}F/N^{3/2}$ extracted from 
fig.~\ref{fig:M_extrapolation} (Right)
is plotted against $\sqrt{k}$.
Our data are in good agreement with 
the result (\ref{M-N3-2}) predicted from eleven-dimensional
supergravity, which is represented by the solid line.
}
\label{fig:ABJM_vs_11dSUGRA}
\end{figure}

In fig.~\ref{fig:ABJM_vs_11dSUGRA} we plot
$h_0 (k)$
against $\sqrt{k}$,
which confirms 
the prediction (\ref{M-N3-2})
from eleven-dimensional supergravity 
for $k=1,2,\cdots,10$ very precisely. 

\subsection{Finite $N$ effects}

\begin{figure}[t]
 \begin{center}
 \includegraphics[width=7.4cm]{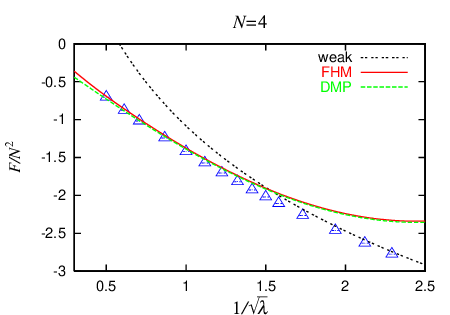}
 \includegraphics[width=7.4cm]{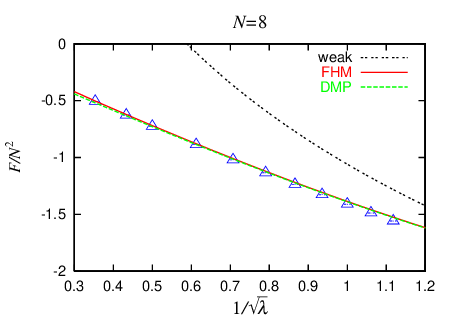}
 \end{center}
 \caption{The free energy of the ABJM theory for $N=4$ (Left)
and $N=8$ (Right) is
plotted against $1/\sqrt{\lambda}$. 
The solid line and the dashed line represent
the FHM result (\ref{FHM}) and the DMP result (\ref{strong}), respectively.
The dotted line represent the perturbative results (\ref{finNsL}). 
}
 \label{fig:free_energy_each_N}
 \end{figure}

\begin{figure}[t]
 \begin{center}
 \includegraphics[width=7.4cm]{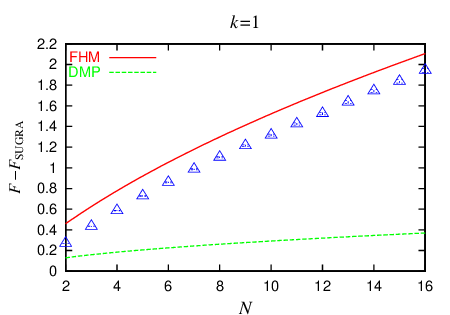}
 \includegraphics[width=7.4cm]{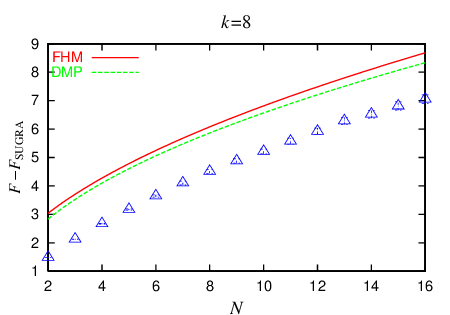}
 \end{center}
 \caption{The free energy of the ABJM theory
after subtraction of the dominant term (\ref{M-N3-2})
is plotted against $N$
for $k=1$ (Left) and $k=8$ (Right).
The solid line and the dashed line represent
the FHM result (\ref{FHM}) and the DMP result (\ref{strong}), respectively.
  }
 \label{fig:free_energy_sugra-diff}
 \end{figure}

One of the important results on finite $N$ effects in the
free energy is that the $1/N$ corrections around the planar limit
are resummed in a closed form (\ref{FHM})
neglecting the worldsheet instanton effect.
In fig.~\ref{fig:free_energy_each_N} we plot our results
for $N=4$ and $N=8$ and compare them with the FHM result (\ref{FHM})
and the DMP result (\ref{strong}).
We find that both FHM and DMP are close to our data at strong coupling,
but the difference between them is too small 
to see whether FHM is doing any better than DMP.
This is simply because the term 
(\ref{M-N3-2}), which commonly exists 
in both results, is dominating over the difference.
We therefore plot $F - F_{\rm SUGRA}$  against $N$
for $k=1$ (Left) and $k=8$ (Right)
in fig.~\ref{fig:free_energy_sugra-diff}.
The leading large-$N$ behavior of the plotted quantity
is
$\frac{\pi}{\sqrt{2}} (\frac{1}{24}+\frac{1}{3k^2} )k^{3/2}\sqrt{N}$
for FHM and 
$\frac{\pi}{24\sqrt{2}} k^{3/2}\sqrt{N}$
for DMP, 
where the difference comes from 
the $\lambda^2 / (3N^2)=1/(3k^2)$ term
in (\ref{lambda_ren}).
The difference becomes negligible for $k=8$, but it is significant for
$k=1$, in which case our data are indeed closer to FHM than to DMP.

\begin{figure}[t]
\begin{center}
\includegraphics[width=10cm]{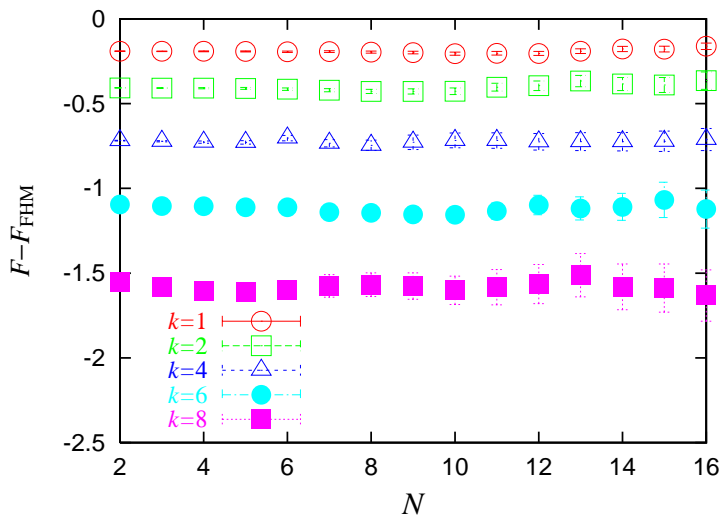}
\end{center}
\caption{
The difference $F-F_{\rm FHM}$ is plotted against $N$
for various values of $k$. 
It reveals non-negligible discrepancies for each $k$,
which are almost independent of $N$.
}
\label{fig:free_energy_fixed_k}
\end{figure}

We also find some discrepancy between our result and FHM,
which are almost independent of $N$.
To see it more directly,
we plot in fig.~\ref{fig:free_energy_fixed_k} 
the difference between our result and the FHM result for various $k$. 
It turns out that
the discrepancies are indeed almost independent of $N$.
This strongly suggests that the FHM result correctly incorporates 
the finite $N$ effects 
except for a term which depends only on $k$.
Note that this discrepancy cannot be explained by
the worldsheet instanton effect ${\rm O}(e^{-2\pi\sqrt{\lambda}})$,
which is neglected in FHM.
While this discrepancy does not affect
the M-theory limit 
corresponding to the strict $N\rightarrow \infty$ limit
for fixed $k$,
it is non-negligible when one considers $1/N$ corrections.
As we will see in section \ref{sec:deviation_origin}, 
this discrepancy coincides with $A(k) -\frac{1}{2}\log{2}$ in eq.~(\ref{fermi}) 
by Fermi gas approach \cite{Marino:2011eh} for small $k$ and with the constant map contribution for all $k$.

\section{Interpretation of the discrepancies}
\label{sec:deviation_origin}
In this section we provide an interpretation of the discrepancies
between our data and the known analytical results,
which we observe in the previous section.

\subsection{Genus expansion}
\label{sec:genus-exp}
Let us consider the planar limit, in which
$g_s N=2\pi i N/k = 2\pi i\lambda$ is kept fixed.  
In that limit, the free energy can be expanded with respect 
to the genus as
\begin{eqnarray}\label{genus}
F(g_s,\lambda)
&=&
\sum_{g=0}^\infty F_g(\lambda)g_s^{2g-2}
\nonumber\\
&=&
-\frac{N^2}{\left(2\pi\lambda \right)^2} F_0(\lambda)
+F_1(\lambda) -\frac{\left( 2\pi\lambda \right)^2}{N^2}
F_2(\lambda)+\cdots  \,.
\end{eqnarray}
Below we consider the free energy order 
by order in this expansion. 

\sss{Planar contribution}

The planar contribution $-k^2 F_0(\lambda)/(4\pi^2 )$
can be studied by the saddle point method,
and the $F_0(\lambda)$ can be determined by 
solving \cite{Marino:2009jd,Drukker:2010nc,Marino:2011nm,Halmagyi:2003ze} 
\begin{eqnarray}\label{dF}
\partial_\lambda F_0(\lambda)
&=&
\frac{\kappa}{4}G^{2,3}_{3,3}
\left(
\begin{array}{c|c}
\begin{matrix}
\frac12&\frac12&\frac12  \\
0&0&-\frac12
\end{matrix}&-\dfrac{\kappa^2}{16}
\end{array}
\right)
+\frac{i\pi^2\kappa}{2}{}_3F_2
\left(\frac12,\frac12,\frac12;1,\frac32;-\frac{\kappa^2}{16}\right)
\,,
\\
\label{lambda}
\lambda(\kappa)
&=&
\frac{\kappa}{8\pi}
{}_3F_2\left(\frac12,\frac12,\frac12;1,\frac32;-\frac{\kappa^2}{16}\right)
\,,
\end{eqnarray}
where $G^{2,3}_{3,3}$ is the Meijer G-function\footnote{The Meijer 
G-function is defined by
\[
G^{m,n}_{p,q}\left(
\begin{array}{c|c}
\begin{matrix}
a_1 & \cdots & a_p\\
b_1 & \cdots & b_q
\end{matrix}& x
\end{array}\right)
= \frac{1}{2\pi i} \int_L \frac{\prod_{j=1}^m 
    \Gamma (b_j -s )     \prod_{j=1}^n    \Gamma (1-a_j +s )}
{\prod_{j=m+1}^q  \Gamma (1-b_j +s ) \prod_{j=n+1}^p \Gamma (a_j -s )}
x^s ds \,,
\]
where the path $L$ is chosen in an appropriate way depending
on the parameters.
See, for instance, ref.~\cite{table_int} for the details.
} 
and ${}_3F_2$ is the hypergeometric function.
Note that these equations are exact for arbitrary $\lambda$,
and hence they fully incorporate the worldsheet instanton effect. 
One can obtain $F_0(\lambda)$ by carrying out the integration over
$\lambda$ as
\begin{eqnarray}
\label{F0-cond}
F_0(\lambda)
=F_0 (0) +\int_0^\lambda d\lambda'\,\partial_{\lambda'}F_0(\lambda')
=F_0 (0) +\int_0^{\kappa(\lambda)}d\kappa'\,\frac{\partial\lambda'}{\partial\kappa'}\,
\partial_{\lambda'}F_0(\lambda') \,.
\end{eqnarray}
At weak coupling $\lambda\ll 1$, in particular, 
one obtains
\ba
F_0(\lambda)
 \,&=& \, 
F_0 (0)+
\tilde{F}_0(\lambda)
+ {\rm O}(\lambda^9) \,, \\
\tilde{F}_0(\lambda)
 \,&=&\, 
2\pi^2\lambda^2\left(3-2\log\frac{\pi\lambda}{2}\right)
+\frac{4\pi^4\lambda^4}{9}-\frac{61\pi^6\lambda^6}{450}
+\frac{12289\pi^8\lambda^8}{79380} \,.
\label{weak-2}
\ea
By comparing this with the perturbative calculation (\ref{weak}),
one finds $F_0(0)=0$.

At strong coupling $\lambda\gg 1$, one obtains
\ba
\label{F0}
F_0(\lambda)
&=&
c_0+ \hat{F}_0(\lambda)
+ {\rm O}\left(e^{-8\pi\sqrt{2\hlam}}\right) \,, \\
\hat{F}_0(\lambda)
&=&
\frac{4\pi^3\sqrt{2}}{3}\hlam^{3/2}
-e^{-2\pi\sqrt{2\hlam}}
+e^{-4\pi\sqrt{2\hlam}}\left(\frac98+\frac{1}{\pi\sqrt{2\hlam}}\right)
\nt&&
-e^{-6\pi\sqrt{2\hlam}}\left(\frac{82}{27}
+\frac{9\sqrt{2}}{4\pi\sqrt{\hlam}}
+\frac{1}{\pi^2\hlam}+\frac{\sqrt{2}}{12\pi^3\hlam^{3/2}}\right) \,,
\ea
where $\hlam =\lambda-1/24$.
Here $c_0$ is an ``integration constant'', 
which has been set zero in the previous works, 
for instance, in ref.~\cite{Drukker:2010nc},
which leads to eq.\,(\ref{strong}).

\begin{figure}[t]
\begin{center}
\includegraphics[width=7.4cm]{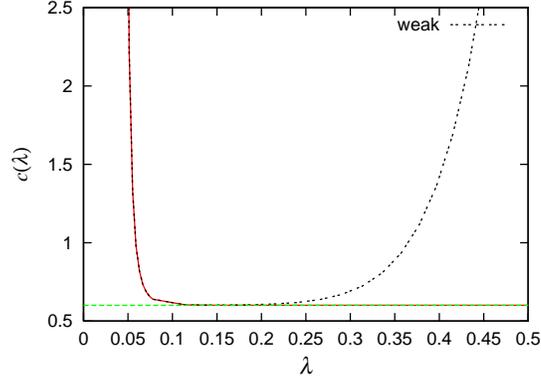}
\end{center}
\caption{The solid line represents
$c(\lambda ) \equiv F_0(\lambda) - \hat{F}_0(\lambda)$,
where $F_0(\lambda)$ is evaluated numerically by
performing the integral (\ref{F0-cond}) explicitly.
The ``integration constant'' 
$c_0$ in (\ref{F0}) is obtained as (\ref{c})
from the asymptotic value of $c(\lambda)$ at large $\lambda$.
The dotted line represents 
$\tilde{F}_0(\lambda)- \hat{F}_0(\lambda)$, where
$\tilde{F}_0(\lambda)$ is the result at weak coupling given by
(\ref{weak-2}). The matching of the weak coupling result and the
strong coupling result around $\lambda \sim 0.15$ also requires 
a similar value of $c_0$ in (\ref{F0}).
}
\label{fig:integration-constant}
\end{figure}

In fig.~\ref{fig:integration-constant}
we plot $c(\lambda) \equiv F_0(\lambda) - \hat{F}_0(\lambda)$,
where $F_0(\lambda)$ is evaluated numerically by
performing the integral (\ref{F0-cond}) explicitly.
As $\lambda$ increases, we
find that $c(\lambda)$ approaches a \emph{nonzero} constant
\begin{eqnarray}\label{c}
c_0 \simeq 0.60103 \,,
\end{eqnarray}
which coincides with $\zeta(3)/2 \simeq 0.601028$
obtained as the constant map contribution at genus zero \cite{HAE,constant_map,Marino:2009dp}.
The value of $a_0$ in (\ref{d_planar}) predicted from the above
calculation is
$a_0 =-c_0/4\pi^2 \simeq -0.015224$, which
agrees well with the discrepancy (\ref{d_planar2}) observed 
in the planar limit.

\sss{Higher genus contributions}

Next we discuss the discrepancy 
at higher genus,
which can be also interpreted as the constant map contributions
as in the planar part. 
Let us note that the analytical results up to the constant map 
have been obtained for genus one and two in 
terms of 
modular forms
as \cite{Drukker:2010nc,Drukker:2011zy,Akemann:1996zr,Huang:2006si}
\ba
\label{F1F2}
F_{\rm modular}^{(1)}(\lambda)&=&
-\log\eta(\tau) \,,  \\
F_{\rm modular}^{(2)} (\lambda)&=&
\frac{1}{432\vth_2^4 \vth_4^8}
\left(-\frac53E_2^3+3\vth_2^4E_2^2-2E_4E_2\right) \nonumber \\
&~& +\frac{16\vth_2^{12}+15\vth_2^8\vth_4^4
+21\vth_2^4\vth_4^8+2\vth_4^{12}}{12960\vth_2^4\vth_4^8}
\,,
\label{F1F2-2}
\ea
where 
$\eta(\tau)$ is the Dedekind eta function, 
$E_n (\tau )$ is the Eisenstein series of weight $n$, 
$\vartheta_n(\tau)$ is the theta function,
and 
$\tau (\lambda )$ is defined as
\ba
\tau (\lambda )=i\frac{K( \sqrt{1+\kappa^2 /16} )}{K( \frac{i\kappa}{4})}
=1+\frac{i}{4\pi^3}\partial^2_\lambda F_0(\lambda ) \,,
\ea
where $K(x)$ is the complete elliptic integral of the first kind.

\begin{figure}[t]
\begin{center}
\includegraphics[width=7.4cm]{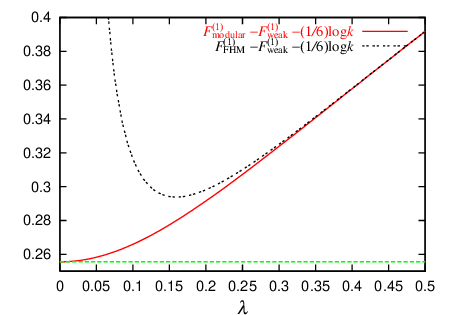}
\includegraphics[width=7.4cm]{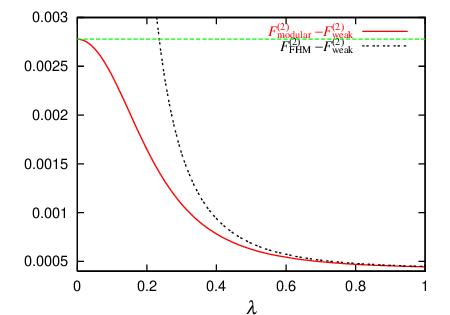}
\end{center}
\caption{(Left) The solid line represents
$F_{\rm modular}^{(1)} -\left( F_{\rm weak}^{(1)} +\frac{1}{6}\log{k}
\right)$, which is  
the difference between the modular expression (\ref{F1F2}) 
and the weak coupling result for genus one.
It approaches a constant smoothly for $\lambda \rightarrow 0$,
which gives $c_1$ in (\ref{c2-val}).
The dotted line represents 
$F_{\rm FHM}^{(1)} -\left( F_{\rm weak}^{(1)} +\frac{1}{6}\log{k}
\right)$, which diverges as $\lambda \rightarrow 0$ since the FHM result
neglects the worldsheet instanton effect.
(Right) The solid line represents
$F_{\rm modular}^{(2)} -F_{\rm weak}^{(2)} $, which is  
the difference between the modular expression (\ref{F1F2-2})  
and the weak coupling result for genus two.
It approaches a constant smoothly for $\lambda \rightarrow 0$,
which gives $c_2$ in (\ref{c2-val}).
The dotted line represents 
$F_{\rm FHM}^{(2)} - F_{\rm weak}^{(2)} $, 
which diverges as $\lambda \rightarrow 0$ since the FHM result
neglects the worldsheet instanton effect.
Here  $F_{\rm FHM}^{(2)}$ is given by
$F_{\rm FHM}^{(2)} =\frac{1}{144\sqrt{2}\:\!\pi}\hlam^{-1/2}-\frac{1}{48\pi^2}\hlam^{-1}+\frac{5}{96\sqrt{2}\:\!\pi^3}\hlam^{-3/2}$.
}
\label{fig:genus12}
\end{figure}

In fig.~\ref{fig:genus12} we plot
$F_{\rm modular}^{(1)} -\left( F_{\rm weak}^{(1)} 
+\frac{1}{6}\log{k}  \right)$ (Left)
and 
$F_{\rm modular}^{(2)}-F_{\rm weak}^{(2)}$ (Right)
against $\lambda$, where we have defined the weak coupling results
\begin{eqnarray}
\label{weak1}
F_{\rm weak}^{(1)}(\lambda)\,&=&\,
-\frac16\left(\log \lambda+\log k\right)+2\zeta'(-1)
\  , \\
F_{\rm weak}^{(2)}(\lambda)\,&=&\,-\frac{B_4}{16\pi^2\lambda^2}
\,, 
\label{weak2}
\end{eqnarray}
at genus one and two, respectively, as 
can be read off from (\ref{finNsL2}). 
We find in both cases that the result approaches a constant 
as $\lambda \rightarrow 0$, which gives
\beq
c_1 \simeq  0.25558 \,,  \quad\quad
c_2 \simeq 0.0027777 \,,
\label{c2-val}
\eeq
respectively.
This suggests that in the weak coupling region
there are additional terms given by
\ba\label{ab}
\Delta F^{(1)}\,=\,-\frac16\log k -c_1 \,, \quad\quad
\Delta F^{(2)}\,=\, -c_2 \,,
\ea
which precisely coincide with the constant map contributions \cite{HAE,constant_map,Marino:2009dp} 
\ba\label{const-map}
&&\back
\text{for genus 1\,:}\qquad\quad\!\!
-\frac16\log k +2\zeta'(-1) +\frac16\log\frac{\pi}{2} \,,
\nt&&\back
\text{for genus $g\geq 2$\,:}\quad
\frac{4^g B_{2g} B_{2g-2}}{4g(2g-2)(2g-2)!}\,.
\ea
Since the FHM result (\ref{FHM}) reproduces the previous results
in the genus expansion,
the FHM result must also have the additional contributions
\begin{eqnarray}
\label{correction_p}
F - F_{\rm FHM} 
\simeq 
-c_0 \frac{k^2}{4\pi^2} -\frac{1}{6}\log{k} - c_1 + 
c_2 \frac{4\pi^2}{k^2} + {\rm O}(k^{-4})\,,
\end{eqnarray}
where the worldsheet instanton effect is neglected.

\begin{figure}[t]
 \begin{center}
 \includegraphics[width=7.4cm]{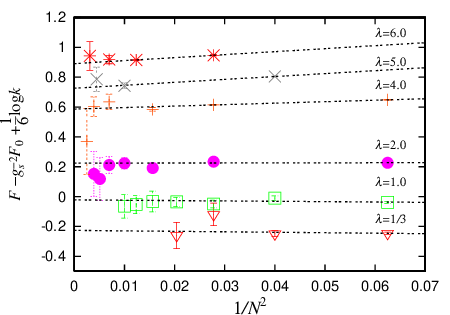}
 \includegraphics[width=7.4cm]{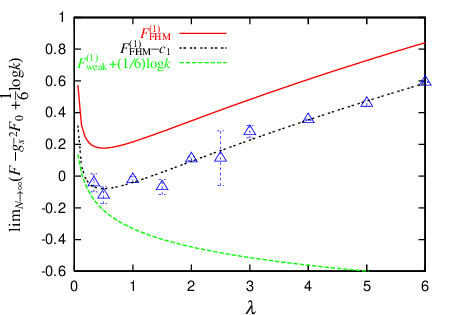}
 \end{center}
 \caption{
(Left) $F-g_s^{-2} F_0 + \frac{1}{6}\log{k}$ is plotted against
$1/N^2$. The results are nicely fitted to straight lines,
which enables us to extract the genus one contribution reliably.
(Right) The genus one contribution extracted from the left panel
is plotted against $\lambda$. The solid line represents the
genus one contribution to $F_{\rm FHM}$, whereas the dotted line
represents $F_{\rm FHM}^{(1)} -c_1$, where $c_1$ is given by (\ref{c2-val}).
The dashed line represents the weak coupling behavior given by
(\ref{weak1}) with the $-\frac{1}{6}\log{k}$ term being subtracted.
}
 \label{fig:genus1_limit}
 \end{figure}

As we did in the case of planar contribution, we can test whether 
the discrepancy in the genus one contribution 
between our data and the previous analytical results
can be explained by the additional terms identified above.
In fig.~\ref{fig:genus1_limit}
we extract the genus one contribution from our data in the following
way.
First we subtract from our data for the free energy, 
the planar contribution
$g_s^{-2} F_0(\lambda)$, where $F_0(\lambda)$ is
obtained by (\ref{F0-cond}),
and subtract also the term $-\frac{1}{6}\log{k}$ that appears in the weak coupling
result (\ref{weak1}).
Then we plot the result after these subtractions against $1/N^2$ in
fig.~\ref{fig:genus1_limit} (Left), 
which can be nicely fitted to straight lines. 
The intercepts give the 
values of the genus one contribution for each $\lambda$,
which are plotted against $\lambda$ in fig.~\ref{fig:genus1_limit}
(Right).
We find that the result is in good agreement with 
the genus one contribution of $F_{\rm FHM}$ after making a constant shift 
by $- c_1$ determined as (\ref{c2-val}).

\subsection{Finite $N$ effects}\label{sec:constant_map_vs_Fermi_gas}

\begin{figure}[t]
\begin{center}
\includegraphics[width=7.4cm]{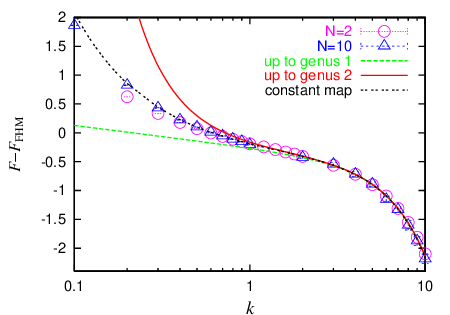}
\includegraphics[width=7.4cm]{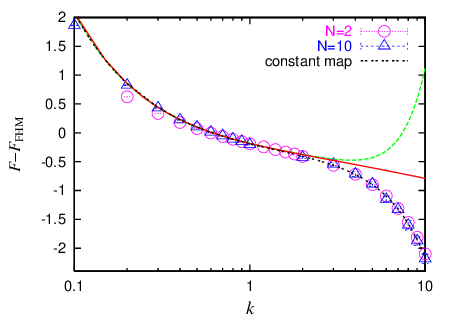}
\end{center}
\caption{
(Left) 
The discrepancy of the free energy between our data and the FHM result
is plotted against $k$, and
compared with the analytical results around the planar limit for $N=2$ and $N=10$ .
The dashed and the solid lines represent the correction (\ref{correction_p}) 
up to genus one term and up to genus two term, respectively.
The dotted line represents the sum of the constant map contributions at all genus (\ref{const-map-sum}).
(Right) The same as the left panel except that 
our data is compared with the result obtained by the Fermi gas approach.
The dotted and the solid lines represent the analytical results
(\ref{fermi}) with $A(k)$ given by
(\ref{Ak-expand}) and (\ref{Ak1}), respectively.
The dotted line represents again
the sum of the constant map contributions at all genus (\ref{const-map-sum}).
} 
\label{fig:large-lambda}
\end{figure}

Let us see how well the FHM result with the corrections
(\ref{correction_p}) does at finite $N$.
In fig.~\ref{fig:large-lambda} we plot $F - F_{\rm FHM}$, i.e.,
the discrepancies between our result and the FHM result
for $N=2$ and $N=10$ against $k$.
At $k\gtrsim 1$, the $N=2$ data and the $N=10$ data are on top of each
other as anticipated from fig.~\ref{fig:free_energy_fixed_k}.
In this regime, the discrepancies are in good agreement with 
the corrections (\ref{correction_p}) identified in 
section \ref{sec:genus-exp}.

It is interesting to see what happens if we go to smaller $k$ region
in fig.~\ref{fig:large-lambda}
although non-integer $k$ is not physical in the original gauge theory.
Firstly we start to see some difference
between $N=2$ and $N=10$, which implies that there is some
$N$ dependence which is not captured by the FHM result
in this regime.
We speculate that the $N$ dependence is due to the membrane instanton effect \cite{Drukker:2011zy,Becker:1995kb},
which behaves as ${\rm O}( e^{- \pi \sqrt{2kN}} )$,
since the worldsheet instanton effect is negligible in this regime.
Secondly, we find that the discrepancy
between our result and the FHM result
no longer agrees with (\ref{correction_p}) including corrections
up to genus two. 
This is understandable since the higher genus terms 
become non-negligible as one goes to 
smaller $k$ (larger $g_s$).

On the other hand, the free energy at small $k$ is calculated 
by the Fermi gas approach as (\ref{fermi}).
We find that our data for $N=10$ interpolate nicely
the behavior (\ref{fermi}) at small $k$
and the behavior (\ref{correction_p}) at large $k$.
This also supports our speculation that the difference
between the $N=2$ and $N=10$ data at small $k$
is due to the membrane instanton effect,
which is 
neglected in the Fermi gas approach.
Note, in particular, that 
the Fermi gas approach yields correction to the FHM result
in odd powers of $k$, whereas the genus expansion yields even powers
of $1/k$.
Our data nicely interpolate the two asymptotic behaviors,
which are smoothly connected around $k\sim 1$.

Finally let us consider the sum of
the constant map contributions at all genus (\ref{const-map}),
which is given by\footnote{An assumption is needed to obtain (\ref{const-map-sum}) and (\ref{cm-2}).  
For details, see appendix \ref{appendix:con_fermi}.}
\ba\label{const-map-sum}
F_{\rm const}
&=&-\frac{\zeta(3)}{8\pi^2}k^2-\frac16\log k+\frac16\log\frac{\pi}{2}
+2\zeta'(-1)\nt
&& -\frac{1}{3} \int_0^\infty dx\ \frac{1}{e^{kx} -1} 
    \left( \frac{3}{x^3} -\frac{1}{x} -\frac{3 }{x \sinh^2{x}}   \right) \,.
\ea
In fig.~\ref{fig:large-lambda} we find that this function 
agrees well with the discrepancy between our result and the FHM result 
for the whole range of $k$ investigated, including $k\lesssim 1$. 
Since the Fermi gas result (\ref{fermi}) also gives accurate description there, 
it is natural to guess that they are actually the same. 
Indeed, as we show in appendix~\ref{appendix:con_fermi}, 
the sum of the constant map (\ref{const-map-sum}) 
can be expanded around $k=0$ as
\ba\label{cm-2}
F_{\rm const}
&=&
\frac{2\zeta(3)}{\pi^2 k}-\frac12\log 2
+\sum_{n=1}^\infty \frac{(-1)^n}{(2n)!}B_{2n} B_{2n-2}\pi^{2n-2}k^{2n-1}
\nt
&=& -\frac{1}{2} \log{2} +\frac{2\zeta (3)}{\pi^2 k} -\frac{k}{12}-\frac{\pi^2 k^3}{4320}
    +\frac{\pi^4 k^5}{907200} +\cdots  \ ,
\ea
which exactly reproduces the result of
the Fermi gas approach (\ref{fermi}) to the $k^3$ term.\footnote{Note
that the Fermi gas calculation has been done only to the $k^3$ term, and
the higher order terms are just an educated guess. Calculations at higher
orders would be an interesting future direction.}
Remarkably, the constant map contribution (\ref{const-map-sum}) and 
the expansion (\ref{cm-2}) are equivalent at any
$k$.
Therefore, we expect that the result (\ref{cm-2}) is the all-order 
form of $A(k)-\frac12\log 2$ in the Fermi gas approach.
In other words, 
we expect that the expansions of the free energy around 
$k=\infty$ (the constant map contribution) 
and $k=0$ (the Fermi gas approach) give the same answer 
after taking sums to all orders.  
In this sense the free energy around the planar and M-theory limits 
are smoothly connected with each other.

\section{Summary and discussions}
\label{sec:con}

In this paper we have established a simple numerical method 
for studying
the ABJM theory on a three sphere for arbitrary rank $N$ 
at arbitrary Chern-Simons level $k$.
The crucial point is that we are able to rewrite the ABJM matrix
model, which is obtained after applying the localization technique,
in such a way that the integrand becomes positive definite.
By using this method, we have confirmed from first principles
that the free energy in the M-theory limit 
grows proportionally to $N^{3/2}$ as predicted from the
eleven-dimensional supergravity.
We have also found that 
the FHM formula with 
the additional terms 
describes the free energy of the ABJM theory
in the type IIA superstring and M-theory regimes. 
Analytic form of the additional terms is discussed in detail, 
and beautiful agreement between planar and M-theory regions is found. 
These additional terms become important when we consider the quantum string
effect in the AdS/CFT duality. 

There are many issues worth being addressed by using our method. 
Most importantly from the conceptual point of view,
we can use the free energy obtained for finite $N$ and finite $k$ 
to test the AdS/CFT duality at the quantum string level. 
At the tree level, or equivalently in the planar limit, 
there is strong evidence that the gauge theory correctly 
describes the string $\alpha'$ effect.
For example, the D0-brane quantum mechanics reproduces 
the $\alpha'$ correction to the black 0-brane solution
in type IIA superstring theory \cite{Hanada:2008ez}.  
However, no definite conclusion is obtained for quantum string
corrections so far. 
In fact, as pointed out in ref.~\cite{Fuji:2011km}, 
the FHM formula does not seem to agree with a prediction from
the string theory side \cite{Bergman:2009zh}. 
This disagreement is not solved even if we take into account
of the corrections found in this paper.
Some of the possible solutions to this puzzle includes:
(i) one has to consider some different gauge theory such as
$SU(N)_k\times SU(N)_{-k}$ theory, which gives different $1/N$ corrections, 
(ii)  one has to refine the argument on the string theory side,
and (iii) the AdS/CFT does not hold at the quantum string level. 
In particular, the scenario (i) can be tested straightforwardly 
by extending our method.

We consider it equally important to study {\it quantum} M-theory.
While there is very little knowledge on it so far,
we may hope to get some insight through intensive numerical studies 
of the ABJM theory.
In fact similar attempts have been made recently \cite{Hanada:2009ne,Hanada:2011fq} 
using the BFSS matrix theory \cite{Banks:1996vh}.
Numerical studies suggest that the prediction from the dual string
theory for the scaling dimension of a certain class of operators continues 
to hold in the M-theory region. 
Similar or possibly more striking features of
M-theory may show up by studying the ABJM theory numerically.

While we have focused on the free energy as the most fundamental
quantity in the ABJM theory, our method can be used
to calculate the expectation values of BPS operators.
For instance, it is possible to calculate the expectation value 
of the circular Wilson loop for various representations. 
They are conjectured to be related to the string worldsheet area and the D-brane world-volume 
in the type IIA superstring region, respectively. It would be interesting 
to test this conjecture and to see the stringy corrections. 

Our method can be also applied to other theories,
which have recently attracted much attention in connection to the F-theorem 
and the entanglement entropy.
For example, one can study the necklace-type quiver 
discussed in ref.~\cite{Gulotta:2011si}.  
We can also study other gauge groups such as $O(N)$ and $USp(N)$
studied in ref.~\cite{Gulotta:2011vp,Gulotta:2012yd}.  
Detailed studies of these theories outside the planar limit, in
particular, would be very interesting. 
For instance, the so-called orbifold equivalence, which is usually proven 
to hold only in the planar limit, can hold outside the planar limit 
in these models \cite{Hanada:2011yz,Hanada:2011zx}.  
Note also that the ABJM matrix model is related to the lens space matrix model, 
which appears in the context of the topological string theory. 
It is therefore conceivable that 
there might be some applications to the topological string theory as well.

We hope that the results of this work are convincing enough
to show the power of the numerical approach,
and that many more applications other than those listed above
would reveal themselves as we proceed further.

\subsection*{Acknowledgment}

The authors would like to thank
N.~Drukker, S.~Hirano, D.~Honda, I.~Kanamori, M.~Mari\~no, Y.~Matsuo, T.~Takimi
and D.~Yokoyama for valuable discussions. 
In particular, we greatly appreciate M.~Mari\~no's important comment 
on the constant map, 
which was crucial in arriving at complete understanding on the correction 
to the FHM formula. 
M.~Hanada, M.~Honda and J.~Nishimura are grateful to the hospitality 
of the Kavli Institute for Theoretical Physics in UCSB,
during the KITP program 
``Novel Numerical Methods for Strongly Coupled Quantum Field Theory
and Quantum Gravity''.
This research was supported in part
by the National Science Foundation under
Grant No.~PHY11-25915.
M.~Honda, Y.~Honma and S.~Shiba are supported by Grant-in-Aid for JSPS fellows (No.~22-2764, 22-3677 and 23-7749).
The work of J.~Nishimura is supported in part by 
Grant-in-Aid for Scientific Research
(No.\ 20540286 and 23244057) from JSPS.
S.~Shiba would like to thank the 
hospitality of theoretical physics group at Imperial College London.
Computations have been carried out on 
PC clusters at KEK Theory Center and B-factory Computer System.

\appendix
\section{Basics and details of the Monte Carlo simulation}
\label{appendix:simulation}
In this section we explain the basics and details of the Monte Carlo
simulation\footnote{There are 
many good references on Monte Carlo methods. See, e.g., ref.~\cite{Rothe:2005nw}.}. 
Let us consider the action 
\begin{eqnarray}
S(N,k;x)
=
-\log\left(\frac{\prod_{i<j}\tanh^2((x_i-x_j)/2k)}{\prod_i 2\cosh(x_i/2)}\right). 
\label{action_real}
\end{eqnarray}
(Below we abbreviate $S(N,k;x)$ as $S(x)$.)
Let ${\cal O}(x)$ be an ``observable'', which is a function of $\{x_i\}$. 
In general, it is difficult to calculate the expectation value of ${\cal O}$ defined by 
\begin{eqnarray}
\langle {\cal O}\rangle
=  \frac{\int d^Nx\,{\cal O}(x)e^{-S(x)}}{\int d^Nx\, e^{-S(x)}} \,.  
\end{eqnarray}
A brute force integration is not practical unless the number of
variables $N$ is very small such as $N\lesssim 5$. 
Monte Carlo simulation is a practical tool, which enables this
calculation even for large $N$.

In Monte Carlo simulation, a series of configurations $\{x_i\}$ 
\begin{eqnarray}
\{x_i^{(0)}\}
\ \to\ 
\{x_i^{(1)}\}
\ \to\ 
\{x_i^{(2)}\}
\ \to\ 
\cdots
\end{eqnarray}
is generated in such a way that 
the probability with which a configuration $\{x_i\}$ appears approaches 
$e^{-S(x)}/Z$ as the number of configurations increases. 
More precisely, we require that 
the probability $w_k(\{x_i\})$ with which 
a configuration $\{x_i\}$ appears at $M$-th step converge to
$e^{-S(x)}/Z$ as
\begin{eqnarray}
\lim_{M\to\infty}
w_M(\{x_i\})
=
\frac{e^{-S(x)}}{Z} \,. 
\end{eqnarray}
Then the expectation value can be obtained 
by simply taking an average over the configurations
$\{x_i^{(n)}\}$ as
\begin{eqnarray}
\langle{\cal O}(x)\rangle
= \lim_{M\to\infty}
\frac{1}{M}\sum_{n=1}^M
{\cal O}(x_i^{(n)}) \,. 
\end{eqnarray}
This can be achieved by generating the series with a transition probability 
$P(\{x_i^{(n)}\}\to\{x_i^{(n+1)}\})$, 
which (neglecting a few technical details) satisfies following conditions.  
\begin{itemize}
\item
{\it Markov chain}. ---
The transition probability from $\{x_i^{(n)}\}$ to $\{x_i^{(n+1)}\}$ 
does not depend on previous configurations $\{x_i^{(l)}\}\ (l<n)$.  

\item
{\it Ergodicity}. ---
For any pair of configurations $\{x\}$ and $\{x'\}$,
there is nonzero transition probability within finite steps. 

\item
{\it Aperiodicity}. ---
The probability from $\{x_i\}$ to $\{x_i\}$ is always nonzero.

\item
{\it Positive state}. ---
All configurations have finite mean recurrence time\footnote{
If $P_{ii}^{(n)}$ is the probability to get from $\{x\}$ to $\{x\}$ in $n$-steps of the Markov chain,
without reaching this configuration at any intermediate step, 
then the mean recurrence time of $\{x\}$ is defined by $\tau_i = \sum_{n=1}^\infty n P_{ii}^{(n)}$. }.

\item
{\it Detailed balance}. --- The following equality should hold for
arbitrary pairs of configurations $\{x\}$ and $\{x'\}$.
\begin{eqnarray}
e^{-S(x)}P(x\to x')
=
e^{-S(x')}P(x'\to x) \,. 
\end{eqnarray}

\end{itemize}

There are many ways to satisfy these conditions. 
In this work we use the Hybrid Monte Carlo (HMC) method \cite{Clark:2003na}. 
We introduce fictitious momentum variables $p_i\ (i=1,2,\cdots,N)$,
which are conjugate to $x_i$, and consider a Hamiltonian 
\begin{eqnarray}
H=\sum_i\frac{p_i^2}{2}+S(x)\,. 
\end{eqnarray} 
Starting with an initial configuration $\{x^{(0)}_i\}$, 
we generate a series of configurations $\{x^{(n)}_i\}\ (n=1,2,\cdots)$ 
by repeating the following steps: 
\begin{itemize}
\item
Generate $p^{(n-1)}_i$ randomly, 
with a probability weight $e^{-\left(p^{(n-1)}_i\right)^2/2}$ \,. 

\item
Starting with a configuration $\{x_i, p_i\}=\{x^{(n-1)}_i,
p^{(n-1)}_i\}$, 
get a new configuration $\{x'_i, p'_i\}$ 
by the ``molecular dynamics'' explained below. 

\item
``Accept'' the new configuration $\{x'_i, p'_i\}$ ({\em i.e.}~take
$\{x^{(n)}_i\}=\{x'_i\}$) 
with a probability $\min\{1,e^{H-H'}\}$, 
where $H$ and $H'$ are the value of the Hamiltonian calculated 
with $\{x_i, p_i\}$ and $\{x'_i, p'_i\}$, respectively.  
When the new configuration is rejected, we keep an old configuration, 
so that $\{x^{(n)}_i\}=\{x_i\}=\{x^{(n-1)}_i\}$. 

\end{itemize}

The ``molecular dynamics'' is defined as follows. 
First we introduce a fictitious time $\tau$ and 
consider the time evolution according to the Hamilton equations
\begin{eqnarray}
\frac{dx_i}{d\tau}=\frac{dp_i}{d\tau}=\frac{\partial H}{\partial
  p_i}=p_i \,, 
\qquad
\frac{dp_i}{d\tau}=-\frac{\partial H}{\partial x_i}=-\frac{\partial
  S}{\partial x_i} \,. 
\end{eqnarray}
If we solve the Hamilton equations exactly, 
the Hamiltonian is conserved. 
In practice, we solve them approximately
by discretizing the differential equations, 
so the Hamiltonian is not conserved exactly.  
We denote the time step as $\Delta\tau$ and the 
number of steps as $N_\tau$. 
Then $x'_i\equiv x_i(N_\tau\Delta\tau)$ and $p'_i\equiv
p_i(N_\tau\Delta\tau)$ are 
calculated by using the following ``leap-frog method'', 
starting with $x_i(0)\equiv x_i$ and $p_i(0)\equiv p_i$.
\begin{itemize}

\item
$x_i \left( \frac{\Delta\tau}{2} \right) =x_i(0)+p_i(0)\cdot \frac{\Delta\tau}{2}$

\item
$p_i(\Delta\tau)=p_i(0)-\left.\frac{\partial S}{\partial x_i}\right|_{\tau=\frac{\Delta\tau}{2}}\cdot\Delta\tau$

\item
$x_i \left( \frac{3}{2}\Delta\tau \right) =
x_i \left( \frac{\Delta\tau}{2} \right) +p_i(\Delta\tau)\cdot \Delta\tau$

\item
$p_i(2\Delta\tau)=p_i(\Delta\tau)-\left.\frac{\partial S}{\partial
  x_i}\right|_{\tau=\frac{3}{2}\Delta\tau}
\cdot\Delta\tau$

\item
$\cdots$

\item
$x_i((N_\tau-1/2)\Delta\tau)=x_i((N_\tau-3/2)\Delta\tau)+p_i((N_\tau-1)\Delta\tau)\cdot \Delta\tau$

\item
$p_i(N_\tau\Delta\tau)=p_i((N_\tau-1)\Delta\tau)-
\left.\frac{\partial S}{\partial x_i}\right|_{\tau=(N_\tau-1/2)\Delta\tau}\cdot\Delta\tau$

\item
$x_i(N_\tau\Delta\tau)=x_i((N_\tau-1/2)\Delta\tau)+p_i((N_\tau)\Delta\tau)\cdot \frac{\Delta\tau}{2}$

\end{itemize}
Note that the leap-frog method is designed so that the reversibility is satisfied.
Namely, by starting with the final configuration $\{x'_i\}$ and
$\{p'_i\}$ and reversing the time, 
the initial configuration $\{x_i\}$ and $\{p_i\}$ is reproduced. 
As a result, the detailed balance condition is satisfied.\footnote{For
  simplicity, 
let us assume $H>H'$. 
(The argument for the case with $H<H'$ is the same.)
Because of the reversibility, the transition probabilities are 
\begin{eqnarray}
P(\{x_i\}\to\{x'_i\})
&=& e^{-p^2/2}/\sqrt{\pi}\times {\rm min}\{1,e^{H-H'}\} \\
&=& e^{-p^2/2}/\sqrt{\pi}
\end{eqnarray}
and
\begin{eqnarray}
P(\{x'_i\}\to\{x_i\})
&=& e^{-p^{\prime 2}/2}/\sqrt{\pi}\times {\rm min}\{1,e^{H'-H}\} \\
&=& e^{-p^2/2+(S(x')-S(x))}/\sqrt{\pi} \,. 
\end{eqnarray}
Therefore, 
\begin{eqnarray}
e^{-S(x)}P(\{x_i\}\to\{x'_i\})
&=& e^{-S(x)}\times
e^{-p^2/2}/\sqrt{\pi} \\
&=& e^{-S(x')}P(\{x'_i\}\to\{x_i\}) \,. 
\end{eqnarray}
}

In the simulation, $N_\tau$ and $\Delta\tau$ should be chosen 
so that a good approximation is achieved with fewer configurations. 
For that purpose, (i) the change at each transition should be
sufficiently large, and 
(ii) the acceptance rate should be large. 
The first condition is achieved by taking $N_\tau\Delta\tau$
sufficiently large. 
However, if we fix $\Delta\tau$ and take larger $N_\tau$, 
the Hamiltonian is not conserved at all, 
and the new configurations are hardly accepted.
Therefore one has to take $\Delta\tau$ smaller 
so that the conservation of the Hamiltonian becomes better. 
In actual simulations (at $N=20$ and $k=5$, for example), 
we took $\Delta\tau\sim 0.1$ and $N_\tau\sim 200$, so that the acceptance rate 
is around 0.8. As an initial configuration, we choose $x_i^{(0)}$ to be a random number in $[-0.5,0.5]$. 
 
In Monte Carlo simulation, 
configurations with larger path-integral weight (``important samples'') appear more often. 
For this reason it is called also the {\it importance sampling}. 
Since the region of configuration space, which gives dominant
contribution
to the path integral is typically quite limited,
a good approximation can be achieved with a rather small number of important samples. 
This should be contrasted to the usual brute force integration, 
in which most of the CPU time is wasted for calculating the integrand
for unimportant configurations.


In Monte Carlo simulation, as we have described above, 
configurations are generated with the 
{\it probability} $e^{-S}/Z$. 
Therefore, the Monte Carlo method works only 
if the path-integral weight $e^{-S}$ is real and positive. 
If the measure $e^{-S}$ is not real and positive, 
the model is said to suffer from the {\it sign problem} or the {\it phase problem}; 
here ``sign'' and ``phase'' mean the negative sign and the complex phase of the integration weight. 
In the original form of the ABJM matrix model \eqref{PF2}, 
the partition function is given by an integration of a complex function. 
Therefore it suffers from the sign problem, and the Monte Carlo method is not applicable.

\section{Derivation of the sign-problem-free form 
of the ABJM matrix model}
\label{appendix:derivation}
In this section we derive the sign-problem-free form of the ABJM matrix model,
which was used in ref.~\cite{Kapustin:2010xq,Okuyama:2011su,Marino:2011eh} 
for a different purpose. 
Let us start with the ABJM matrix model (\ref{PF2}).
We are going to use the Cauchy identity\footnote{See the 
appendix of ref.~\cite{Kapustin:2010xq} for the proof of this identity.}
\begin{equation}
\frac{\prod_{i<j} (u_i -u_j ) (v_i -v_j) } {\prod_{i,j} (u_i +v_j ) }
= \sum_\sigma (-1)^\sigma \prod_i \frac{1}{u_i +v_{\sigma (i)}} \,.
\end{equation}
Here $\sigma$ runs through all permutations.
By setting $u_i =e^{\mu_i} ,v_i =e^{\nu_i}$, it becomes
\begin{eqnarray*}
\frac{\prod_{i<j} (e^{\mu_i} -e^{\mu_j} ) (e^{\nu_i} -e^{\nu_j}) } {\prod_{i,j} (e^{\mu_i} +e^{\nu_j} ) }
&=& \sum_\sigma (-1)^\sigma \prod_i \frac{1}{e^{\mu_i} +e^{\nu_{\sigma
      (i)}} }  \,. 
\end{eqnarray*}
From this, we obtain
\begin{eqnarray*}
\frac{\prod_{i<j}\Bigl[ 2\sinh{\left( \frac{\mu_i -\mu_j}{2} \right) } \Bigr] 
                     \Bigl[ 2\sinh{\left( \frac{\nu_i -\nu_j}{2} \right) } \Bigr] }
         {\prod_{i,j} \Bigl[ 2\cosh{\left( \frac{\mu_i -\nu_j}{2} \right) } \Bigr] }
&=& \sum_\sigma (-1)^\sigma \prod_i 
     \frac{1}{ 2\cosh{\left( \frac{\mu_i -\nu_{\sigma (i)}}{2} \right)
     }  }  \,.
\end{eqnarray*}
Therefore, the partition function can be written as
\begin{eqnarray*}
&& Z(N,k) \\
&=& \frac{1}{N!^2} \sum_{\sigma ,\sigma^\prime} (-1)^{\sigma +\sigma^\prime} 
    \int \frac{d^N \mu}{(2\pi )^N} \frac{d^N \nu}{(2\pi )^N} 
    \prod_i   \frac{1}{ \Bigl[ 2\cosh{\left( \frac{\mu_i -\nu_{\sigma (i)}}{2} \right)  } \Bigr]
                        \Bigl[ 2\cosh{\left( \frac{\mu_i -\nu_{\sigma^\prime (i)}}{2} \right)  } \Bigr]  } 
       \exp{\Bigl[ \frac{ik}{4\pi}\sum_{i=1}^N (\mu_i^2 -\nu_i^2 )    \Bigr] }  \\
&=& \frac{1}{N!} \sum_\sigma  (-1)^\sigma  
    \int \frac{d^N \mu}{(2\pi )^N} \frac{d^N \nu}{(2\pi )^N} 
    \prod_i   \frac{1}{ \Bigl[ 2\cosh{\left( \frac{\mu_i -\nu_i}{2} \right)  } \Bigr]
                        \Bigl[ 2\cosh{\left( \frac{\mu_i -\nu_{\sigma (i)}}{2} \right)  } \Bigr]  } 
       \exp{\Bigl[ \frac{ik}{4\pi}\sum_{i=1}^N (\mu_i^2 -\nu_i^2 )
           \Bigr] }  \,.
\end{eqnarray*}
By using the formula
\[
\frac{1}{2\cosh{p}} = \frac{1}{\pi} \int dx \frac{e^{\frac{2i}{\pi}p
    x}}{2\cosh{x}} \,,
\]
we obtain
\begin{eqnarray*}
&&  \sum_\sigma  (-1)^\sigma
    \prod_i   \frac{1}{ \Bigl[ 2\cosh{\left( \frac{\mu_i -\nu_i}{2} \right)  } \Bigr]
                        \Bigl[ 2\cosh{\left( \frac{\mu_i -\nu_{\sigma (i)}}{2} \right)  } \Bigr]  }  \\
&=& \sum_\sigma  (-1)^\sigma \frac{1}{\pi^{2N}} \int d^N x d^N y\ 
  \frac{\exp{\Bigl[ \frac{i}{\pi}\sum_i (\mu_i -\nu_i )x_i
                   +\frac{i}{\pi}\sum_i (\mu_i -\nu_{\sigma(i )})y_i   \Bigr]}}
        {\prod_i 2\cosh{x_i} \cdot 2\cosh{y_i} } \\
&=& \sum_\sigma  (-1)^\sigma \frac{1}{\pi^{2N}} \int d^N x d^N y\ 
  \frac{\exp{\Bigl[ \frac{i}{\pi}\sum_i (\mu_i -\nu_i )x_i
                   +\frac{i}{\pi}\sum_i (\mu_i y_i  -\nu_i y_{\sigma (i)})   \Bigr]}}
        {\prod_i 2\cosh{x_i} \cdot 2\cosh{y_i} } \,.
\end{eqnarray*}
Therefore, the partition function becomes\footnote{In the Fermi gas approach \cite{Marino:2011eh}, 
the integrand is identified with a partition function for the ideal
Fermi gas given by
\begin{eqnarray}
Z(N,k)
&=& \frac{1}{N!} \sum_\sigma  (-1)^\sigma    
            \int d^N x\ \prod_{i=1}^N \rho (x_i , x_{\sigma (i)} ) \,,
\label{fermi_wave}
\end{eqnarray}
where $\rho (x_1 , x_2 )$ is interpreted as the 
one-particle density matrix
\begin{equation}
\rho (x_1 ,x_2 ) 
= \frac{1}{2\pi k} \frac{1}{\left( 2\cosh{\frac{x_1}{2}}\right)^{1/2}} 
\frac{1}{\left( 2\cosh{\frac{x_2}{2}}\right)^{1/2}}
   \frac{1}{2\cosh{\left( \frac{x_1 -x_2}{2k}\right)}} \,.
\end{equation}
}
\begin{eqnarray}
Z(N,k)
&=& \frac{1}{N!} \sum_\sigma  (-1)^\sigma   \frac{1}{\pi^{2N}} \int
d^N x d^N y\   
\frac{1}{\prod_i 2\cosh{x_i} \cdot 2\cosh{y_i} } \nonumber \\
&&   \int \frac{d^N \mu}{(2\pi )^N} \frac{d^N \nu}{(2\pi )^N} 
\exp{\Bigl[ \frac{i}{\pi}\sum_i (\mu_i -\nu_i )x_i
           +\frac{i}{\pi}\sum_i (\mu_i y_i  -\nu_i y_{\sigma (i)})  
+\frac{ik}{4\pi}\sum_{i=1}^N (\mu_i^2 -\nu_i^2 )  \Bigr]} \nonumber \\
&=& \frac{1}{N!} \sum_\sigma  (-1)^\sigma   \frac{1}{\pi^{2N}} 
\int d^N x d^N y\   \frac{1}{\prod_i 2\cosh{x_i} \cdot 2\cosh{y_i} } 
   \int \frac{d^N \mu}{(2\pi )^N} \frac{d^N \nu}{(2\pi )^N} \nonumber \\
&& \exp{\Bigl[ \frac{ik}{4\pi}\sum_{i=1}^N \left( \mu_i +\frac{2}{k}(x_i +y_i)\right)^2
              -\frac{ik}{4\pi}\sum_{i=1}^N \left( \nu_i 
+\frac{2}{k}(x_i +y_{\sigma (i)})\right)^2  \Bigr]} \nonumber \\
&& \exp{\Bigl[ -\frac{i}{k\pi}\sum_{i=1}^N \left( (x_i +y_i )^2 -(x_i
       +y_{\sigma (i)})^2  \right) \Bigr] } 
\nonumber \\
&=& \frac{1}{N!} \sum_\sigma  (-1)^\sigma   \frac{1}{\pi^{2N}} \int
d^N x d^N y\   
\frac{1}{\prod_i 2\cosh{x_i} \cdot 2\cosh{y_i} } 
   \int \frac{d^N \mu}{(2\pi )^N} \frac{d^N \nu}{(2\pi )^N} \nonumber \\
&& \exp{\Bigl[ \frac{ik}{4\pi}\sum_{i=1}^N \mu_i^2 -\frac{ik}{4\pi}\sum_{i=1}^N \nu_i^2    
               -\frac{2i}{k\pi}\sum_{i=1}^N x_i (y_i -y_{\sigma (i)}) \Bigr] } \nonumber \\
&=& \frac{1}{N!} \sum_\sigma  (-1)^\sigma   \frac{1}{k^N \pi^{2N}} 
             \int d^N x d^N y\   \frac{1}{\prod_i 2\cosh{x_i} \cdot 2\cosh{y_i} } 
       e^{ -\frac{2i}{k\pi}\sum_{i=1}^N x_i (y_i -y_{\sigma (i)})  } \nonumber \\
&=& \frac{1}{N!} \sum_\sigma  (-1)^\sigma   \frac{1}{(k \pi )^N} 
             \int d^N y\   \frac{1}{\prod_i 2\cosh{\left( 
\frac{y_i -y_{\sigma (i)}}{k} \right)} \cdot 2\cosh{y_i} } \nonumber \\
&=& \frac{1}{N!} \sum_\sigma  (-1)^\sigma    
             \int \frac{d^N x}{(2\pi k)^N}\ 
            \frac{1}{\prod_i 2\cosh{\left( \frac{x_i}{2}\right) }
\cdot 2\cosh{\left( \frac{x_i -x_{\sigma (i)}}{2k} \right)}   } \,.
\label{grand_cano}
\end{eqnarray}

We use the Cauchy identity again:
\[
\sum_\sigma (-1)^\sigma \prod_i 
     \frac{1}{ 2\cosh{\left( \frac{x_i -x_{\sigma (i)}}{2k} \right) }  } 
=    \frac{\prod_{i<j} \Bigl[ 2\sinh{\left( \frac{x_i -x_j}{2k} \right)} \Bigr]^2 }
          {\prod_{i,j} \Bigl[ 2\cosh{\left( \frac{x_i -x_j}{2k} \right)} \Bigr]}
=    \frac{1}{2^N} \prod_{i<j} \tanh{^2\left( \frac{x_i -x_j}{2k}
  \right) } \,.
\] 
Thus we arrive at the final expression
\begin{eqnarray}
Z(N,k)
&=& \frac{1}{2^N N!}  \int \frac{d^N x}{(2\pi k)^N}\ 
         \frac{\prod_{i<j} \tanh{^2\left( \frac{x_i -x_j}{2k} \right) }}
              {\prod_i 2\cosh{\left( \frac{x_i}{2}\right) }} \,, 
\label{cauchy_ABJM}
\end{eqnarray}
which does not have a sign problem.

\section{The relation between the constant map and the Fermi gas result}
\label{appendix:con_fermi}
In this appendix we show the correspondence between the constant map contribution and 
the Fermi gas result $A(k)-\frac{1}{2}\log{2}$,
which is derived by the large-$k$ and small-$k$ expansions, respectively.

As we mentioned earlier, the constant map contribution $F_{\rm const}$ is given by 
\begin{\eq}\label{F_const}
F_{\rm const} = \sum_{g=0}^\infty g_s^{2g-2} F_{\rm const}^{(g)} \ ,
\end{\eq}
where the coefficients $F_{\rm const}^{(g)}$ are  
\begin{\eqa}
F_{\rm const}^{(0)} = \frac{\zeta (3)}{2},\quad 
F_{\rm const}^{(1)} = 2\zeta^{\prime}(-1)
+\frac{1}{6}\log{\frac{\pi}{2k}} \ , \quad 
F_{\rm const}^{(g\geq 2)} = 4^g \frac{B_{2g}
  B_{2g-2}}{(4g)(2g-2)(2g-2)!} \ .
\end{\eqa}
In order to evaluate the summation more easily,  
we use the integral representation of the Bernoulli number, 
\begin{\eq}
B_{2g} = (-1)^{g-1} 4g \int_0^\infty \frac{x^{2g-1}}{e^{2\pi x} -1} dx 
\quad (g=1,2,\cdots )  \ .
\label{bernoulli_int}
\end{\eq}
By using this representation, we obtain
\begin{\eqa}
\sum_{g=2}^\infty g_s^{2g-2} F_{\rm const} ^{(g)}
&=& \sum_{g=2}^\infty (-1)^{g-2} g_s^{2g-2} 4^g \frac{B_{2g} }{(2g)(2g-2)!} 
  \int_0^\infty \frac{x^{2g-3}}{e^{2\pi x} -1} dx  \NN \\
&=& g_s^{-2}\int_0^\infty dx\ \frac{x^{-3}}{e^{2\pi x} -1} 
   \sum_{g=2}^\infty (-1)^{g}  \frac{B_{2g} }{(2g)(2g-2)!} (2g_s
   x)^{2g}   \ .
\end{\eqa}
This summation is easily performed by using a formula
\begin{\eq}
\sum_{g=2}^\infty  (-1)^g \frac{B_{2g}}{(2g)(2g-2)!} (2z)^{2g}
= \frac{1}{3}\left( 3+z^2 -\frac{3z^2}{\sin^2{z}}   \right) \ .
\end{\eq}
Note that the series converges only for $|z|=|g_s x|<\pi$.
However, since the result  of the summation in the right hand side is well-defined even for $|g_s x|\geq \pi$,
we analytically continue $g_s x$ to the whole region including $|g_s x|\geq \pi$
and assume that this does not affect the result of the integration\footnote{
Even if this assumption is not valid, the discrepancy at small $k$ is expected to be $e^{-2\pi/k}$ at most. 
}.

By substituting $z=g_s x$ in the formula, the summation is rewritten as
\begin{\eqa}
\sum_{g=2}^\infty g_s^{2g-2} F_{\rm const}^{(g)} 
&=& -\frac{k^2}{12\pi^2} \int_0^\infty dx\ \frac{x^{-3}}{e^{2\pi x} -1} 
    \left( 3 -\frac{4\pi^2}{k^2} x^2 
-\frac{12\pi^2 x^2}{k^2 \sinh^2{(\frac{2\pi x}{k})}}   \right)  \ .
\end{\eqa}
By changing the variable as $t=\frac{2\pi x}{k}$, we obtain a simpler form, 
\begin{\eqa}
\sum_{g=2}^\infty g_s^{2g-2} F_{\rm const}^{(g)} 
&=& -\frac{1}{3} \int_0^\infty dt\ \frac{t^{-3}}{e^{kt} -1} 
    \left( 3 -t^2 -\frac{3 t^2}{ \sinh^2{t}}   \right)  \ .
\label{integral_CF}
\end{\eqa}
Although each term of the integrand is divergent at $t\sim 0$, 
this is canceled with each other, and therefore the integral gives a finite value.
In order to make our analysis easier, 
we will apply the zeta-function regularization to the integral.
  
For later convenience, we decompose the integral as  
\begin{\eqa}
\sum_{g=2}^\infty g_s^{2g-2} F_{\rm const}^{(g)} 
&=& \int_0^\infty dt\ \frac{1}{e^{kt} -1} 
    \left( -\frac{1}{t^3} +\frac{1}{3t}   \right) 
   +\frac{1}{k} \int_0^\infty dt\ \frac{kt}{e^{kt} -1} \frac{1}{t^2
     \sinh^2{t}} \ .
   \label{constant_map_sum}
\end{\eqa}
Note that the first factor in the second term is 
the generating function of the Bernoulli number
\begin{\eq}
\frac{x}{e^{x}-1} =\sum_{n=0}^\infty B_n \frac{x^n}{n!} .
\label{Bernoulli_bokansu}
\end{\eq}
Although this series also converges only for $|x|<2\pi$,
we analytically continue it to the whole region and 
assume that this does not affect the result. 
Then by using the formula
\begin{\eq}
\frac{1}{e^{x}-1} = 
\sum_{m=1}^\infty e^{-mx}\,,\quad \frac{1}{\sinh^2{x}}
=-\sum_{m=1}^\infty m e^{-mx} \,,
\end{\eq}
the integral rewritten as
\begin{\eqa}
\sum_{g=2}^\infty g_s^{2g-2} F_{\rm const}^{(g)} 
&=& \sum_{m=1}^\infty \int_0^\infty dt\ e^{-mkt}  \left( -\frac{1}{t^3} +\frac{1}{3t}   \right)  
     +\frac{4}{k} \sum_{n=0}^\infty \frac{B_n k^n}{n!} \sum_{m=1}^\infty m 
       \int_0^\infty dt\ t^{-2+n}e^{-2mt} \ . \NN \\
\end{\eqa}

The first integral is easily performed by using 
\begin{\eqa}
\int_0^\infty dt\ e^{-st}t^{-1} = -\gamma -\log{s} ,\quad 
\int_0^\infty dt\ e^{-st}t^{-3} =
-\frac{1}{4}s^2 (-3+2\gamma +2\log{s}) \ , \NN
\end{\eqa}
where $\gamma$ is the Euler-Mascheroni constant. 
We obtain
\begin{\eqa}
&& \sum_{m=1}^\infty  \int_0^\infty dt\ e^{-mkt} \left( -\frac{1}{t^3} +\frac{1}{3t}   \right) \NN \\
&=& \frac{k^2}{4} \sum_{m=1}^\infty n^2 (-3+2\gamma +2\log{(km)}) 
   +\frac{1}{3}\sum_{m=1}^\infty (-\gamma -\log{(mk)} ) \NN \\
&=& \frac{k^2}{4}\left[ (-3+2\gamma +2\log{k})\zeta{(-2)} -2\zeta^\prime (-2)  \right] 
    +\frac{1}{3}\left[ (-\gamma -\log{k} )\zeta (0) +\zeta^\prime (0) \right]  \NN \\
&=& -\frac{k^2}{2} \zeta^\prime (-2)  +\frac{1}{6} \left[ (\gamma +\log{k} ) -\log{(2\pi )} \right] . 
\end{\eqa}

Next we evaluate the second integral 
\[
\sum_{m=1}^\infty m 
       \int_0^\infty dt\ t^{-2+n}e^{-2mt} \ . 
\]

\begin{itemize}
\item For $n=0$\\
By using the formula
\begin{\eqa}
\int_0^\infty dt\ e^{-st}t^{-2}
&=& s( -1 +\gamma +\log{s}) \ ,
\end{\eqa}
we obtain
\begin{\eqa}
\sum_{m=1}^\infty m   \int_0^\infty dt\ t^{-2}e^{-2mt} 
&=&  -2\zeta^\prime (-2) \ .
\end{\eqa}

\item For $n=1$\\
\begin{\eqa}
\sum_{m=1}^\infty m   \int_0^\infty dt\ t^{-1}e^{-2mt} 
&=& \frac{1}{12}(\gamma +\log{2}) +\zeta^\prime (-1) \ .
\end{\eqa}

\item For $n\geq 2$\\
By using the formula
\begin{\eqa}
\int_0^\infty dt\ e^{-st}t^{\lambda -1} &=& \Gamma (\lambda )
\frac{1}{s^\lambda}\quad (Re(\lambda ) >0) \ ,
\end{\eqa}
the integral becomes
\begin{\eqa}
\sum_{m=1}^\infty m   \int_0^\infty dt\ t^{-2+n}e^{-2mt} 
&=& \frac{\Gamma (n-1)}{2^{n-1}}\zeta (n-2) \ .
\end{\eqa}

\end{itemize}
Thus the constant map contribution is rewritten as
\begin{\eqa}
F_{\rm const}
&=& -\frac{\zeta (3)}{8\pi^2} k^2 -\frac{1}{6}\log{k} +\frac{1}{6}\log{\frac{\pi}{2}} +2\zeta^\prime (-1) 
    -\frac{k^2}{2} \zeta^\prime (-2) 
     +\frac{1}{6}\left( \gamma +\log{k} -\log{(2\pi )} \right) \NN\\
&&  +\frac{4}{k}\Biggl[ -2 B_0 \zeta^\prime (-2) 
                        +B_1 k \left( \frac{1}{12}(\gamma +\log{2}) +\zeta^\prime (-1)  \right)
                        +\sum_{n=1}^\infty \frac{B_{2n}}{(2n)!}k^{2n} \frac{\Gamma (2n-1)}{2^{2n-1}}\zeta (2n-2) 
   \Biggr] \NN \\
&=& -\left( \frac{\zeta (3)}{8\pi^2} +\frac{\zeta^\prime (-2)}{2} \right) k^2 
     +\left( 2(1+2B_1 )\zeta^\prime (-1) +\frac{1}{6}(1+2B_1 )\gamma +\frac{1}{3}(-1 +B_1 ) \log{2}  \right)  \NN \\
&& -\frac{8}{k}B_0 \zeta^\prime (-2) 
   +\sum_{n=1}^\infty \frac{B_{2n}}{(2n)(2n-1)2^{2n-3}}\zeta
   (2n-2)k^{2n-1} \ . 
\end{\eqa}
Since $B_0 =1$, $B_1 =-\frac{1}{2}$, $\zeta^\prime (-2)=-\frac{\zeta (3)}{4\pi^2}$ and
$\zeta (2n) =(-1)^{n-1} \frac{2^{2n-1}\pi^{2n}}{(2n)!}B_{2n}$, we obtain
\begin{\eqa}
F_{\rm const}
&=& -\frac{1}{2} \log{2}    +\frac{2\zeta (3)}{\pi^2 k} 
   +\sum_{n=1}^\infty (-1)^n \frac{B_{2n}B_{2n-2}}{(2n)!}\pi^{2n-2} k^{2n-1} \label{fermi_all} \\
&=& -\frac{1}{2} \log{2} +\frac{2\zeta (3)}{\pi^2 k} -\frac{k}{12}-\frac{\pi^2 k^3}{4320}
    +\frac{\pi^4 k^5}{907200} +\cdots  \ ,
\end{\eqa}
which is same as $A(k)-\frac{1}{2}\log{2}$ 
derived by the Fermi gas approach \cite{Marino:2011eh} up to the order of ${\rm O}(k^5)$.
Therefore, we expect that this is the all-order form in the Fermi gas picture if we calculate higher order of $k$.

Thus we conclude that 
the constant map contribution and the term $A(k)-\frac{1}{2}\log{2}$ 
in the Fermi gas result
are the series expansions of the integral representation (\ref{integral_CF}) 
around $k=\infty$ and $k=0$, respectively, with the radius of convergence being finite.
In other words, the two expansions are smoothly connected with each other 
by analytic continuation.


\end{document}